\documentclass[preprint,12pt]{elsarticle}
\usepackage{graphicx}
\usepackage{xcolor}
\usepackage{amsmath,amssymb}
\usepackage{float}
\usepackage{physics}
\usepackage{amsfonts}
\usepackage{verbatim}
\usepackage{flushend}
\usepackage{balance}
\usepackage{endnotes}
\usepackage{footnote}
\usepackage{adjustbox}
\usepackage{gensymb}
\usepackage{subfigure}
\usepackage{float}
\usepackage{epigraph}
\usepackage{hhline}
\usepackage{xcolor}
\usepackage[utf8]{inputenc}
\usepackage{setspace}
\usepackage{hyperref}
\usepackage{algorithm2e}
\newcommand{\beq}{\begin{eqnarray}}
\newcommand{\eeq}{\end{eqnarray}}
\newcommand{\dlp}{DL$\_$POLY}

\SetArgSty{textnormal}

\newcommand{\brac}[1]{\left(#1\right)}
\newcommand{\sqbrac}[1]{\left[#1\right]}
\newcommand{\intd}[1]{\text{d}#1}
\newcommand{\tdbyd}[2]{\frac{\text{d}#1}{\text{d}#2}}

\usepackage{amsmath}

\usepackage[normalem]{ulem}

\journal{Computer Physics Communications}

\begin{document}

\title{\dlp{} 5: Calculation of system properties on the fly for very large systems via massive parallelism}

\author{H. L. Devereux$^{1}$, C. Cockrell$^{1,2}$,  A. M. Elena$^{3}$, Ian Bush$^{4,6}$, Aidan B. G. Chalk$^{5}$, Jim Madge$^{3,7}$, Ivan Scivetti$^{3}$, J. S. Wilkins$^{4,6}$, I. T. Todorov$^{1,3}$, W. Smith$^{3}$, K. Trachenko$^{1}$}
\address{$^1$ School of Physical and Chemical Sciences, Queen Mary University of London, Mile End Road, London, E1 4NS, UK}
\address{$^2$ Nuclear Futures Institute, Bangor University, Bangor, LL57 1UT, UK}
\address{$^3$ Scientific Computing Department, Science and Technology Facilities Council, Daresbury Laboratory, Keckwick Lane, Daresbury, WA4 4AD, UK}
\address{$^4$ Scientific Computing Department, Science and Technology Facilities Council, Rutherford Appleton Laboratory, UK -  current address}
\address{$^5$ Hartree Centre, Science and Technology Facilities Council, Daresbury Laboratory, Keckwick Lane, Daresbury, WA4 4AD, UK}
\address{$^6$ Oxford eResearch Centre, University of Oxford - work carried out}
\address{$^7$ The Alan Turing Institute, British Library, 96 Euston Road, London, NW1 2DB - current address}
\begin{abstract}
Modelling has become a third distinct line of scientific enquiry, alongside experiments and theory. Molecular dynamics (MD) simulations serve to interpret, predict and guide experiments and to test and develop theories. A major limiting factor of MD simulations is system size and in particular the difficulty in handling, storing and processing trajectories of very large systems. This limitation has become significant as the need to simulate large system sizes of the order of billions of atoms and beyond has been steadily growing. Examples include interface phenomena, composite materials, biomaterials, melting, nucleation, atomic transport, adhesion, radiation damage and fracture. More generally, accessing new length and energy scales often brings qualitatively new science, but this has currently reached a bottleneck in MD simulations due to the traditional methods of storing and post-processing trajectory files. To address this challenge, we propose a new paradigm of running MD simulations: instead of storing and post-processing trajectory files, we calculate key system properties on-the-fly. Here, we discuss the implementation of this idea and on-the-fly calculation of key system properties in the general-purpose MD code, \dlp{}. We discuss code development, new capabilities and the calculation of these properties, including correlation functions, viscosity, thermal conductivity and elastic constants. We give examples of these on-the-fly calculations in very large systems. Our developments offer a new way to run MD simulations of large systems efficiently in the future. 
\end{abstract}
\maketitle
\noindent \textbf{PROGRAM SUMMARY/NEW VERSION PROGRAM SUMMARY}
  %Delete as appropriate.
\begin{small}
\noindent
{\em Program Title:} \dlp{}\_5                                          \\
{\em CPC Library link to program files:} (to be added by Technical Editor) \\
{\em Developer's repository link:} \url{https://gitlab.com/ccp5/dl-poly}\\
{\em Licensing provisions:} L-GPL v3.0\\
{\em Programming language:} Fortran 2008                                 \\
{\em Supplementary material:}                                 \\
  % Fill in if necessary, otherwise leave out.
% {\em Journal reference of previous version:} Todorov, I.T., Smith, W., Trachenko, K. and Dove, M.T., 2006. DL\_POLY\_3: new dimensions in molecular dynamics simulations via massive parallelism. Journal of Materials Chemistry, 16(20), pp.1911-1918.                 \\
  %Only required for a New Version summary, otherwise leave out.
% {\em Does the new version supersede the previous version?:} Yes   \\
%   %Only required for a New Version summary, otherwise leave out.
% {\em Reasons for the new version:} Development of on-the-fly correlation analysis.\\
  %Only required for a New Version summary, otherwise leave out.
% {\em Summary of revisions:} General code modularisation, a new control format, SPME re-implementation for per-particle contributions, and a general on-the-fly correlation framework.\\
  %Only required for a New Version summary, otherwise leave out.
{\em Nature of problem:} Molecular dynamics is utilised for modelling in many domains including physics, chemistry, biology, materials science, and their interfaces. These applications often call for large scale simulations targeting high-fidelity timescales and ever larger length scales. In all these cases efficient use of computation, storage, and input/output (I/O) load handling on both the software and hardware sides is required to facilitate analysis.\\
{\em Solution method:}
DL\_POLY provides an efficient set of algorithms for molecular simulation alongside a domain-decomposition strategy to distributed computation efficiently across massively parallel CPU systems. These include parallel I/O handling, link-cells, smooth particle mesh Ewald electrostatics, and now a general purpose on-the-fly correlation framework. This latter development addresses the problem of storing and analysing large trajectory files by iteratively computing correlations at runtime.
\\
  %Describe the method solution here.
%{\em Additional comments including restrictions and unusual features (approx. 50-250 words):}\\
  %Provide any additional comments here.
   \\
\end{small}

\section{Introduction}

For a few decades now, molecular simulations have grown into a third distinct line of scientific enquiry in condensed matter, alongside experiment and theory \cite{allentild,frenkel2002understanding}. Molecular dynamics (MD) simulations give coordinates and momenta of simulated particles as a function of time and thus provide the system's trajectory in phase space. This enables us to calculate most important system properties including those related to structure, statistics, dynamics, transport and so on.

MD results are used in a number of ways to provide insight into experiments, from interpretation to prediction and guiding new ones, as well as to test theories and provide insight at the atomic level resolution for materials, scenarios or processes not yet synthesised by researchers or constructed by experimentalists. MD is particularly useful in fields where measurements are inaccessible in experiment, where empirical theory of phenomena at the larger engineering scales can be examined at short time and length scales, such as damage in materials \cite{nordlund2002computational}. Another popular use is extracting statistics and analytics that are impossible to capture by experiment such as chem-informatics analysis (formation of hydrogen bond networks, hydrophobicity, $\pi$-$\pi$ stacking, etc.), the dynamics of defects in materials (defects and faults in crystalline matrices, amorphisation processes) as well as coarse-grained modelling using augmented force-fields of complex dynamics of, for example, proteins. Generally, MD results are also used to generate synthetic benchmark data for fortifying machine-learnt (ML) models.

This interaction between experiments, theory and simulation has proven to be very productive and has naturally found wide applications in physics, chemistry, materials and earth sciences, engineering, biological and medical sciences etc, from purely academic research to real-world industrial applications. These applications include those in the strategic priority areas of energy, environment, advanced materials and health. It is hard to imagine the current state of these areas without MD and other types of computer modelling. Many practising scientists have taken MD simulations on board to the extent that they consider themselves as combined experimentalist-modellers or theorist-modellers.  

As for any method, MD simulations have their limitations and associated domain of applicability. Several important ones have been relaxed or lifted through the developments of the last decades; others remain unsolved, including the impact of small system size on on simulated properties, such as diffusion coefficients \cite{celebi2021finite}.

One of the strengths of MD is its ability to access small length scales of nanometers and short time scales of picoseconds or beyond where few experimental methods can measure. Early MD simulations were able to simulate about 100 particles for typically tens of picoseconds. With increasing computer power, simulations of millions of atoms on a personal computer is now routine, with simulation times extending to nanoseconds and above.

The appetite of the modelling community to simulate larger systems is constantly growing. In many areas, large systems are necessary to reach the required length or energy scales and match these to experiments and relevant physical processes. These areas include interface phenomena, composite materials, biomaterials, melting, nucleation, atomic transport, adhesion, radiation damage and fracture. 

\begin{figure}
    \centering
    \includegraphics[width=\columnwidth]{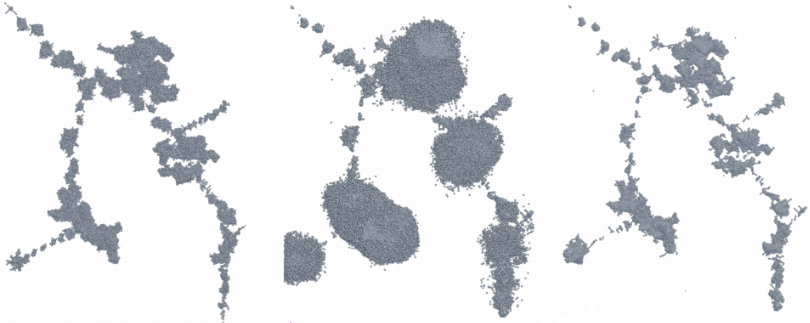}
    \caption{A representative 0.5 MeV collision cascade from our earlier work \cite{Zarkadoula_2013} showing displaced atoms at 0.2, 1.5 ps and 100 ps. The system size is a cube of 2000 \AA\ size with 0.1 billion atoms. Animations showing the propagation of the collision cascade can be watched from \cite{cascade}.}
    \label{casc}
\end{figure}

About a decade ago, we simulated system sizes approaching 1 billion atoms with realistic many-body potentials for iron, using over 60,000 parallel processors \cite{Zarkadoula_2013}. This large system size was required in order to contain radiation damage due to high-energy (in MeV) recoil energy. A representative picture of the damage is shown in Fig. \ref{casc}. We note that the system size, in the order of micrometers, is in interesting proximity to the length scales visible through an optical microscope; simulations and experiments met at this length-scale for the first time.

We found that the nature of the damage and defects in these structures were new and could not be extrapolated from lower-energy events. This had a clear value from scientific perspective and showed that, as is often the case in physics, exploring new length- and energy-scales brings about new effects \cite{Zarkadoula_2013}.

Running these very large simulations, not only brought scientific insights, but also computational insights highlighting the scale of the data. In single precision, the uncompressed positions of 100 million atoms for one timestep costs approximately 1.2GB of disc storage. Computed velocity and force data brings this to 3.6GB. In an MD simulation, certain structural and dynamical properties (for example, viscosity or velocity correlation functions, momentum currents, and radial distribution functions) often need over 10,000-100,000 configurations to be computed, requiring usually two orders of magnitude more timesteps in order to get good-quality data. For such analysis the data requirement reaches 100's of Terabytes. On the other hand, the work file system of a high-performance facility such as UK's ARCHER2 service \cite{archer2} has a total of 10.2 PB \cite{archer-storage} to be shared between all users  (a maximum of 60 TB allocated to a single project or team). Going beyond to 1 billion atoms will require Petabytes of data and yet simulations of this magnitude are increasingly being performed \cite{nguyen2021billion, singharoy2019atoms, dommer2023covidisairborne, stevens2023molecular}. The magnitude of these storage requirements presents a significant storage challenge which must be addressed before any post-processing to obtain results.

Second, a high-performance computing facility such as ARCHER2 spends a significant amount of its time writing the data rather than actually carrying out the numerical simulation, which is an inefficient way of utilising the HPC resource. Even with efficient parallelisation of writing data to storage, a best outcome means that writing a single sample configuration costs as much in CPU time as a timestep.  In the case of 100 million atoms run on 65,000 processors, this fraction was in excess of 5\% for a full production run, including reading and writing.  This implies that each simulation run involving writing the trajectory file, even if feasible in terms of file size, wastes considerable compute power.  Furthermore, the larger the simulation (in model size and time evolution) and the larger the count of processors used to run it, the larger the waste.

In addition to these problems many MD use-cases rely on the analysis of rare events such as protein folding or phase transitions. Both accessing these events and the analysis of them in MD exacerbates the issue of trajectory storage. In this domain the exploitation of collective variables (CV) \cite{fiorin2013using} both to inform sampling or otherwise bias dynamics and to detect or drive these events is well used. As a result software such as PLUMED \cite{bonomi2009plumed, tribello2014plumed, plumed2019promoting} and A4MD \cite{caino2023runtime} act as \textit{in situ} monitors of MD simulations to ingest data trajectory data at runtime to monitor CVs and act upon them.

The problems however extend more generally and will either prevent most users from running and analysing MD simulations of large systems or make these simulations inefficient. In general this challenge can be partially addressed with multiple methodologies related to data compression. Plain ASCII is often a default and the simplest schemes merely reduce written significant figures (for example this is optional in \dlp, and also with GROMACS's XTC format \cite{abraham2015gromacs}). Nevertheless, this comes at potentially significant accuracy losses. Alternatively common lossless formats such as GZIP are also offered, for example in LAMMPS \cite{thompson2022lammps} which can achieve significant file size reductions. However more advanced algorithms specifically for floating-point data exists such as ZFP \cite{lindstrom2014fixed}. This compression can be performed at a relatively low cost including into efficient domain specific data-layouts such as H5MD \cite{de2014h5md} that utilises HDF5 \cite{HDF5} for MD data. For example lossless compression ratios of around 1.44 with ZFP compared with 1.09 for GZIP on the same test data suite can be achieved. With lossy compression this rises up to 10.98 or 226.3 with 32-bit or 16-bit precision. ZFPs throughput rates of around 20 MB/s to 60 MB/s for 64 and 32 bit precision on single CPU cores makes this a negligible cost \cite{lindstrom2014fixed}. In the lossless case this can be expected to result in a 30\% reduction in I/O overheads. If 32-bit (or less) lossy precision is acceptable using ZFP compressed trajectory output and on-the-fly methods will likely perform similarly in terms of workflow core hours. Since the data foot-print will be reduce by around 90\% (or more). Note however that the on-the-fly method involves no precision loss by bit-per-float reduction or lossy compression. It may also be used in conjunction with a more manageable lossy trajectory write-out. For extremely large systems lossless storage and post-processing analysis are still problematic.

Alternatively these problems can be addressed by bypassing trajectory storage entirely. In the simplest cases observables can trivially be computed each step, on-the-fly, without (or whilst) saving the trajectory such as the stress tensor. A more complex example is our radiation damage case study - we overcame these problems by implementing a special on-the-fly algorithm in the general-purpose MD package, \dlp{} \cite{diver2020evolution, dlpoly-ref}. This algorithm specifically finds and analyses displaced and defect atoms in the very large billion-atom system. This was a non-trivial task as the algorithm had to be made consistent with the domain-decomposition infrastructure of \dlp{} \cite{todorov2004dl_poly_3, todorov2006dl_poly_3}. These simulations have provided a useful case study and a motivation for the development of a more general code, with a view to benefit a wider modelling community and to approach the problems in a general way.

Recall that some properties such as energy can be sampled and averaged at each time-step, and these are calculated by \dlp{} and other popular MD codes as a standard output. However, there are other important properties relying on time correlation functions which need to be calculated over the entire trajectory and which are currently calculated by writing out the trajectory file and analysing it \textit{a posteriori}. One common example is the velocity autocorrelation function related to the phonon density of states and power spectrum of the system. Other examples include \textbf{k}-space densities and currents\cite{balucani}. For very large system sizes, the problems discussed above make performing these analyses from trajectory files impractical.

Instead of attempting to save a trajectory and subsequently analysing it, we have enabled the calculation of important physical properties on-the-fly. This sort of on-the-fly calculation will be crucial to the execution and analysis of ever-larger MD simulations, particularly as there is a growing need to simulate larger systems, length and energy scales, coupled with the availability of scalable MD codes and massive parallel computing facilities.

Supported by the EPSRC funding, we have extended the UK flagship MD code, \dlp{}, to calculate system properties on-the-fly culminating in the \dlp{} 5 code available at \cite{dlpoly-code}. \dlp{} is a general-purpose massively parallel MD simulation package that uses a highly efficient set of methods and algorithms, including domain decomposition, linked cells, Daresbury Advanced Fourier Transform, Trotter derived Velocity Verlet integration and RATTLE \cite{dlpoly-ref}. Written to support academic and industrial research, \dlp{} has a wide range of applications and can run on a wide range of computers: from single processor workstations to massively parallel high-performance facilities. During development a strong emphasis was placed on efficient utilization of multi-processor power by optimising memory workload and distribution. This makes it possible to efficiently simulate large systems of billions of atoms. Simulation of systems of this size has been tested, and we are not aware of reasons preventing from running larger systems.  \dlp{} 5 is hosted on GitLab \cite{dlpoly-code} has been a free and open source project (GPLv3.0) since 2020.

In this paper, we document and detail the developments of \dlp{} related to on-the-fly calculation of some key system properties. This included general and substantial refactoring across \dlp{}'s modules to enable the new developments. We begin by discussing the theory and implementation of on-the-fly calculations of correlation functions. We then show results from simulations of small and large scales - up to 100 million atoms - for calculating different properties including viscosity and thermal conductivity, elastic constants, currents, and rigid body properties. These results are compared with experimental results and previous MD simulations. After showing the case studies we discuss benchmarks of our work alongside the costs and benefits or using on-the-fly methods in \dlp, and more generally. We finish with an overview of \dlp{}'s recent history since our last published work \cite{dlpoly-ref}, the 4th and now 5th major releases. Including in particular the refactoring which laid the groundwork for the on-the-fly developments.

\section{System properties on-the-fly}
\subsection{On-the-fly methods}
Transforming data as it is acquired is a paradigm often termed ``online algorithms'' \cite{karp1992line}. The basic idea is that, instead of processing an entire data set a once, an on-the-fly calculation processes the data incrementally, often as the data are generated, received, or simply by processing a data set in chunks. This can incur performance advantages by eliminating the requirement to store the full data set on disc or in memory (RAM). Alternatively input/output (I/O) penalties can be significant when reading from disk, and RAM may be limited. In some cases, the full data set may be either too large to store in memory, or perhaps even unbounded. One example of unbounded data is the internet, where a continuous stream of ``big''-data is encountered on a daily basis \cite{krempl2014open}. In these applications a sliding-window technique is often used where statistics and other data transformations are defined in reference to a fixed-size window \cite{datar2002maintaining} which could be temporal or spatial.

In MD, the phase space configuration of atoms is mapped through simulation time-steps into a trajectory. In order to calculate any properties defined by these time-linked configurations the trajectory may be stored on disc, and an algorithm applied. For example storing atom velocities and calculating the velocity auto-correlation function (VAF). For small systems, it may be possible to store these data at every step, obtaining a VAF accurate to the time-step $\dd t$. As the system size scales this becomes infeasible both in terms of storage and I/O penalties if storing to disc. By conceptualising the simulation trajectory as a big-data stream, we can apply the same ideas of online algorithms to calculate system properties on-the-fly. For modest systems, this presents a convenience. However for large systems, this methodology can give access to properties which otherwise require infeasible storage requirements or some accuracy trade-off.

Multiple system properties depend upon the calculation of correlation functions. As an incomplete list: (1) The VAF is one example which can in turn be used to calculate the vibrational density of states. (2) From Green-Kubo theory \cite{Zwanzig1965, tuckerman2023statistical} the thermal conductivity and viscosity can be calculated by using correlation functions of the heat flux and stress tensor respectively \cite{Zwanzig1965, tuckerman2023statistical}. (3) Elastic constants may also be calculated using either strain correlations \cite{Ray1985Statistical} or stress correlations \cite{squire1969isothermal, lutsko1989generalized}. (4) \textbf{k}-space density correlations may be used to determine thermal-conductivity \cite{cheng2020computing}, and momentum current correlations may be used to calculate structure factors and liquid phonon spectra \cite{balucani, yangprl}.

Motivated by these examples, and others, in \dlp{} we implemented a general on-the-fly correlation algorithm, so that we are able to calculate these values at runtime; that is as the system is simulated, without storing trajectory data to disc.
\subsection{Correlation functions at runtime}
Given two observable quantities $X(s\Delta t)$ and $Y(s\Delta t)$ in MD, where $s\in 1,2, \ldots S$ are simulation time-steps for a time-step of $\Delta t$, we can evaluate the correlation function of these two values at time-lags $l=1, 2, \ldots, L \ll S$ as \cite{tuckerman2023statistical},
\begin{equation}
    C_{XY}(l\Delta t) = \frac{1}{S-l}\sum_{s'=1}^{S-l}X(s'\Delta t)Y((s'+l)\Delta t) \label{eq:cor}
\end{equation}
for one trajectory. Typically statistics are accumulated over multiple trajectories. In any case, directly evaluating equation \ref{eq:cor} is cumbersome as $S$ and $L$ increase. Even more so if the observable is averaged over atoms of molecules (for example, in the case of the VAF). Therefore, often the discrete Fourier transforms of $X$ and $Y$ are taken to access the correlation function using the inverse transform of their product \cite{tuckerman2023statistical}. Whether the sum is taken directly or a Fourier transform method is used $X$ and $Y$ must be stored in memory or on disc for the computation.

We use the multiple-tau correlation algorithm \cite{Ramirez2010Time} within \dlp{} to calculate correlations on-the-fly. The method works by maintaining an hierarchical set of blocks which are updated during the simulation. The blocks store observed data (velocities, stresses, etc.) in a decreasing resolution. The first block stores the observed data intact whilst lower blocks store block averages, computed over the previous level. This is controlled by three parameters, the number of blocks $b$, the number of points within each block $p$, and the size of the block average $m < p$. See Appendix \ref{app:cor} for algorithmic details of our implementation based upon \cite{Ramirez2010Time}.

This approach is similar to the velocity correlator introduced by \citet{frenkel2002understanding} with the additional separation of $p$ (data points stored at each level) and $m$ the averaging parameter. By increasing $b$ and $p$ (with $m$ fixed), later correlation times may be collected at a fixed accuracy. Increasing $m$ has the effect of calculating much longer correlation times for a trade off in accuracy and increased performance. The maximum lag-time of a correlation given these parameters is defined by $(p-1) m^l u\Delta t$ where $\Delta t$ is the simulation time-step and the frequency the correlation is updated is $u$. The trade off in performance terms scales as $\mathcal{O}(p\frac{m+1}{m})$. The actual performance will also depend on the values correlated (e.g. velocity across all atoms, or system stress) and other system options (such as electrostatics).

In \dlp{} we support correlating key quantities such as components of heat-flux, velocity, the stress tensor, rigid body positions, velocites, and angular velocities, as well as ${\bf k}$-space resolved densities, heat-fluxes, currents, stress tensors, and energy currents. Additionally common statistical values reported by \dlp{}'s statistics module including, but not limited to volume, kinetic/potential energy, and temperature. We do this by defining each as an \texttt{observable}. Quantities that are \texttt{observable} may be juxtaposed in the \texttt{CONTROL} file to define arbitrary correlations. For example a velocity-velocity correlation may be defined by \texttt{v\_x-v\_x} where \texttt{\_x} indicates the component correlated, and \texttt{stress\_xy-stress\_xy} defines a stress correlation of the $xy$-component of the stress tensor. Scalar quantities such as volume can be requested using e.g. \texttt{volume-volume}. The power of the juxtaposition is that any pair of observables can be correlated. 

For each correlation, independently, we also support specifying the parameters $b$, $p$, and $m$ separately, and provide the ability to set the frequency each correlation is updated separately from the frequency of other statistics and output in \dlp{}. This enables high-resolution correlations to be calculated without additional I/O overhead, and without forcing low-resolution correlations to be calculated at unnecessarily high time-resolutions. Differences in correlation frequency are handled in the correlation output by automatically accounting for the different effective time-units in the outputted lag times.

During the simulation, correlators are stored indexed by the unique names of the observables they correlate, i.e. at any point in the simulation, we may trivially obtain the current correlation value for say the heat-flux in $x$, \texttt{heat\_flux\_x-heat\_flux\_x}. Not only does this serve to maintain uniqueness, but also allows us to easily report these values, or process them further, at runtime. The result of each correlator is reported in a new file \texttt{COR} which reports the value of each correlation function as well as potentially a set of derived quantities, at user specified intervals. These derived quantities include viscosity, thermal conductivity, and elastic constants as detailed in the next sections.

\subsection{Viscosity and thermal conductivity}
\label{therm-con}
The viscosity and thermal conductivity are important properties characterising liquids, for example in the evaluation of the performance of molten salts as nuclear reactor coolant fluids, since this is determined by their transport and thermal diffusion characteristics \cite{Rosenthal1970, FRANDSEN2020}; these can be calculated using Green-Kubo theory from correlations of shear-stress and heat-flux respectively \cite{Zwanzig1965, tuckerman2023statistical, zhang2015reliable}.

The shear-viscosity may be calculated using the integral
\begin{equation}
    \eta=\frac{V}{k_{\mathrm{B}}T}\int_0^\infty \dd t \langle \sigma_{xy}(t) \ \sigma_{xy}(0) \rangle, \label{eq:visc}
\end{equation}
where $\langle \cdot \rangle$ is the ensemble average, $T$ is the system temperature and $k_\mathrm{B}$ is Boltzmann's constant. Finally the stress tensor, for pairwise additive potential $U$, is defined as
\begin{equation}
    \sigma_{\alpha\beta} = \frac{1}{V}\sum_{i}m^{i}v^{i}_{\alpha}v^{i}_{\beta} - \frac{1}{V}\sum_{i, j\neq i}\frac{\partial U(r^{ij})}{\partial r}\frac{r_{\alpha}^{ij}r_{\beta}^{ij}}{r^{ij}}, \label{eq:stress}
\end{equation}
where $m^{i}$, ${\bf v}^{i}$, ${\bf r}^{i}$ are particle $i$'s mass, velocity, and position, and ${\bf r}^{ij} = {\bf r}^{i} - {\bf r}^{j}$, and $V$ is the system volume. Finally, $\alpha$ and $\beta$ are orthogonal Cartesian coordinates.

Similarly the thermal conductivity $\kappa$ is related to the heat flux by
\begin{equation}
    \kappa =\frac{V}{3k_{\mathrm{B}} T^2} \int_0^\infty \dd t \langle \mathbf{q}(t)\cdot\mathbf{q}(0)\rangle. \label{eq:therm}
\end{equation}
Where the heat flux is 
\begin{equation}
    \mathbf{q} = \frac{1}{V} \sum_{i=1}^N \biggl( E^{i}{\bf v}^{i}+\frac{1}{2}\sum_{j\neq i}{\bf f}^{ij}\cdot {\bf v}^{i}{\bf r}^{ij}\biggr),
\end{equation}
where $E^{i}$ is the total energy (kinetic plus potential) of particle $i$, and ${\bf f}^{ij}$ is the force exerted on $i$ due to $j$. In Eq. \ref{eq:therm} the dot product has the function of averaging over the dimensions $x$, $y$, and $z$.

For multi-component systems with significant asymmetry (e.g. mass difference) it is necessary to correct for mass flux effects when calculating $\kappa$ in order to compare to experimental data \cite{armstrong2014thermal}. For example the effect is significant in LiF but not in KCl \cite{cockrell2024thermal} which have mass ratios $\frac{6.941}{18.99}=0.37$ and $\frac{35.453}{39.098} = 0.91$ respectively. This can be done by forming multiple Green-Kubo formulae using the heat flux ${\bf q}$ and partial momentum densities for particular species,
\begin{align}
    {\bf j}_{I_{s}} = \frac{1}{V}\sum_{i\in I_{s}}m_{i}{\bf v}^{i}(t),
\end{align}
where $I_{s}$ is the set of atom indices for atoms of species $s$. In \dlp{} the partial momentum densities may be calculated, and correlated, for a selection of species by the user. The values themselves may also be written to the \texttt{HEATFLUX} file alongside ${\bf q}$. In general it is possible to take advantage of momentum conservation in order to not calculate all partial momentum densities. A user is then able to form the necessary correlations for the corrected Green-Kubo formula required.

\subsubsection{Comparison to experiment}
\begin{figure}
    \centering
    \includegraphics[width=\columnwidth]{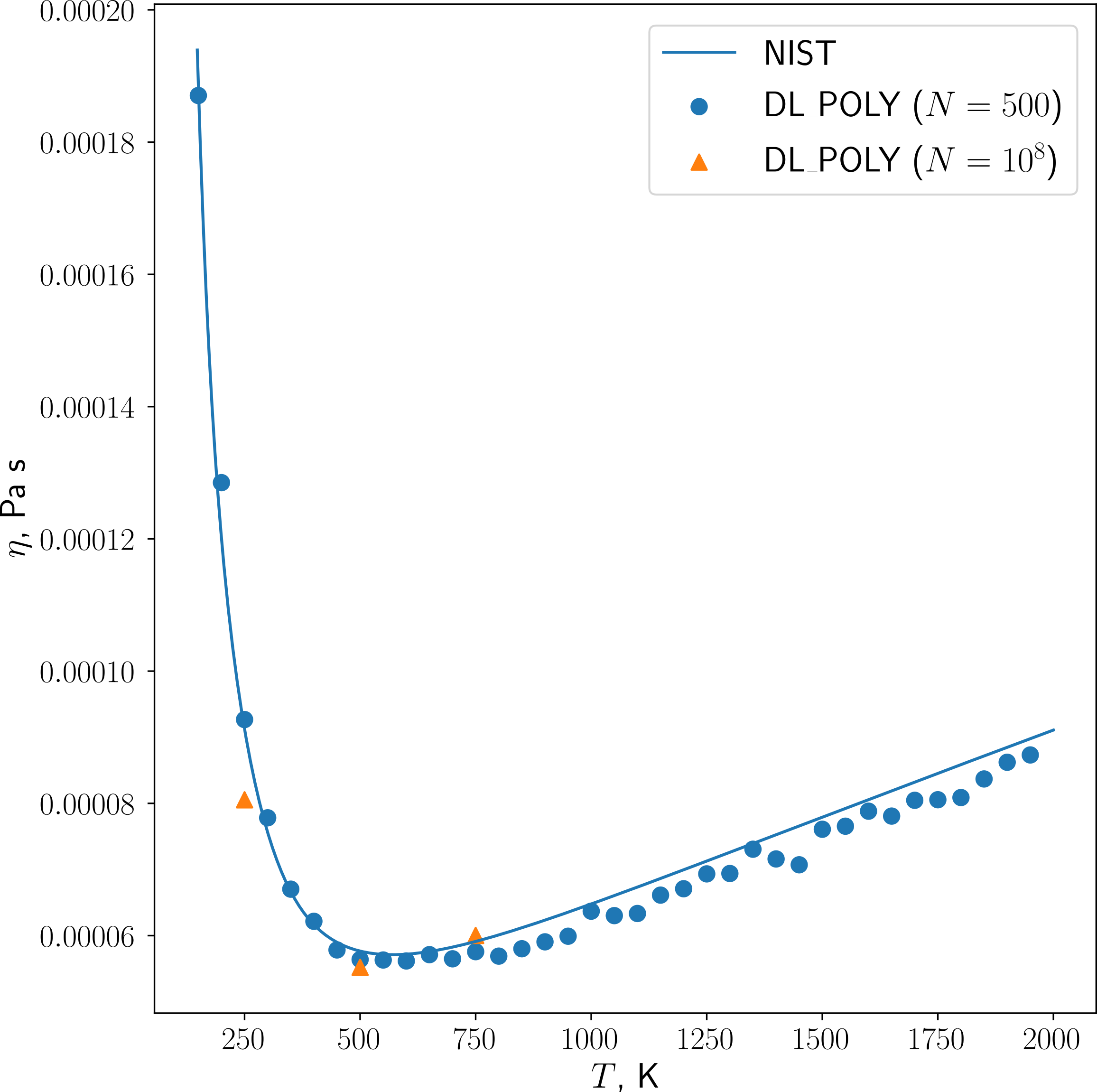}
    \caption{Viscosity for supercritical Argon, as calculated using \dlp{}'s on-the-fly correlations. In each case experimental data from NIST is compared to data from $N=500$ atom simulations averaged over $20$ simulations, and large scale simulations with $N=10^{8}$ atoms averaged over $5$ simulations, also compared to NIST \cite{nist}}
    \label{fig:viscosity}
\end{figure}
The correlation functions integrated in equations \ref{eq:visc} and \ref{eq:therm} can be calculated during an MD run using \dlp{}'s on-the-fly correlator. In each case we can approximate the integral by computing the required correlation to some maximum time-lag $T$ and performing a numerical integration of the correlation functions' values. In each case, when multiple congruent correlation are present (e.g. $xy$ and $zx$ shear stresses), \dlp{} automatically calculates the derived properties (viscosity) listing them for each correlation and also as an averaged quantity. Care must be taken to collect adequate statistics for accurate calculations of these derived quantities. For example \citet{zhang2015reliable} detail a method for choosing an optimal cutoff lag-time.

For argon, there is experimental data for $\kappa$ and $\eta$ at a variety of pressure and temperatures compiled by the National Institute of Standards and Technology (NIST) \cite{nist}. \dlp{} has been used successfully before to calculated viscosity for argon by directly correlating the outputted stress tensor \cite{Cockrell2021Universal}. Here we report new data including thermal conductivity using \dlp{}'s on-the-fly correlator. 

For all simulation work, we began by equilibrating for $10^5$ steps in an NPT ensemble, followed by $10^4$ steps with an NVE ensemble to account for changing thermostats, and then production runs, where statistics are calculated, ran for $10^6$ time-steps. In all cases the time-step was $0.001$ ps. For correlations we collect data for lag times up to $5$ ps. For the small systems ($N=500$) we collect data on independently seeded simulations for $20$ replicates to ensure adequate averaging of the Green-Kubo results. For the large-scale systems, we were able to simulate $5$ replicates for each data point.

\begin{figure}
    \centering
    \includegraphics[width=\columnwidth]{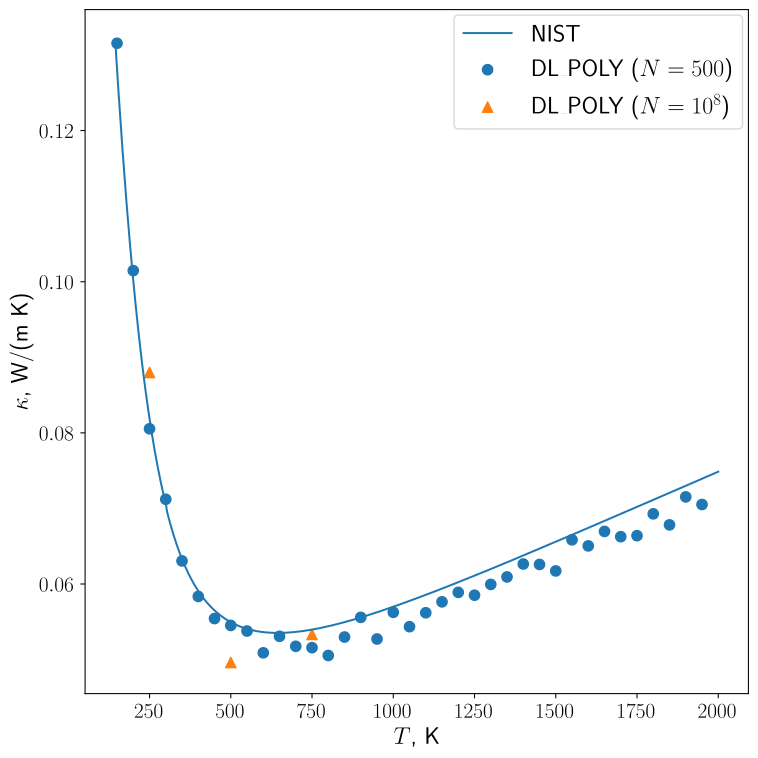}
    \caption{Thermal conductivity for supercritical Argon, as calculated using \dlp{}'s on-the-fly correlations. In each case experimental data from NIST is compared to data from $N=500$ atom simulations averaged over $20$ simulations, and large scale simulations with $N=10^{8}$ atoms averaged over $5$ simulations, also compared to NIST \cite{nist}. See also figure \ref{fig:viscosity}.}
    \label{fig:thermal}
\end{figure}

Figure \ref{fig:viscosity} shows the small- and large-scale system results for viscosity (see Eq. \ref{eq:visc}). We find good agreement with the NIST data as expected \cite{Cockrell2021Universal}. Additionally we find similarly consistent results for the large-scale simulations. Likewise for thermal conductivity, Eq. \ref{eq:therm}, we also find the calculated results are consistent with the experimental data (see Figure \ref{fig:thermal}). The thermal conductivity data notably deviates more, this is due to increased noise between different replicates. For viscosity, we find standard deviations, across the different simulation results, of the order $10^{-6}$ Pa for both system sizes, whereas for thermal conductivity the smaller systems' standard deviations are of the order $10^{-3}$ but for the large-scale systems this rose to $10^{-2}$ at $250$ K and $10^{-3}$ at $500$ K and $750$ K.

Both viscosity and thermal conductivity show an interesting feature: the minima which bound these properties from below. The minima are due to the dynamical crossover of particle dynamics \cite{Cockrell2021Universal}. Interestingly, the values at the minima are fixed by fundamental physical constants such as the Planck constant and electron mass \cite{sciadv,pt2021,momentumprb}.

\subsection{Elastic constants}
\begin{figure}
    \centering
    \includegraphics[width=\columnwidth]{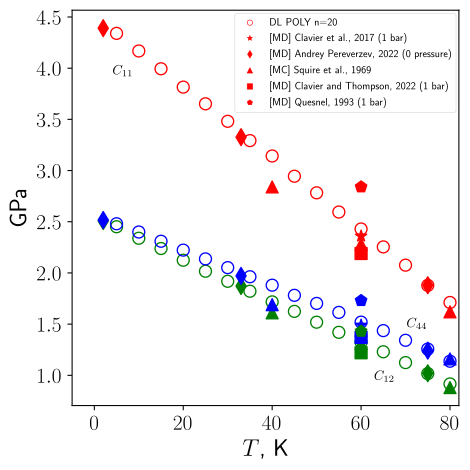}
    \caption{Elastic constants for FCC Argon, as calculated by \dlp{} compared to previous simulation results. In general there is good agreement with data that exists, notably existing data is quite sparse. Our data is averaged over $n=20$ initial conditions, and statistics collected over $T=10^6$ steps, with a $10$ ps maximum correlation lag time.}
    \label{fig:sims-comp}
\end{figure}
Another use case for the on-the-fly correlator is the calculation of elastic constants. In general, the elastic constants of a system are defined by the elements of the elasticity tensor $C_{\alpha\beta\mu\nu}$. The values of which can be calculated, in full, using either the stress-fluctuation of strain-fluctuation methods, or by examining system stress after applying a set of strains. The stress-fluctuation method has been found to be more reliable and faster converging \cite{clavier2017computation, quesnel1993elastic, pereverzev2022isothermal, thompson2022general}. 

For the calculation of elastic constants, we use this method where the elastic constants are given by,
\begin{align}
    C_{\alpha\beta\mu\nu} &= \langle C^{B}_{\alpha\beta\mu\nu}\rangle \nonumber\\&- \frac{V}{k_{\mathrm{B}}T}[\langle \sigma_{\alpha\beta} \sigma_{\alpha\beta}\rangle-\langle \sigma_{\alpha\beta}\rangle\langle \sigma_{\mu\nu}\rangle] \nonumber\\&+ \frac{2Nk_{\mathrm{B}}T}{V}(\delta_{\alpha\mu} \delta_{\beta\nu}+\delta_{\alpha\nu}\delta_{\beta\mu}). \label{eq:elasticity} 
\end{align}
where the parameters $N$ is the atom count, $\delta_{ij}$ is the Kronecker delta, and $\sigma_{\alpha\beta}$ is the microscopic stress tensor as in equation \ref{eq:stress}. Finally the Born term is defined as,
\begin{align}
    C^{B}_{\alpha\beta\mu\nu} = \frac{1}{V}\sum_{i, j \neq i}\biggl(&\frac{\partial^{2} U(r^{ij})}{\partial {r^{ij}}^{2}}\nonumber\\&-\frac{1}{r^{ij}}\frac{\partial U(r^{ij})}{\partial r^{ij}}\biggr)\frac{r^{ij}_{\alpha} r^{ij}_{\beta} r^{ij}_{\mu} r^{ij}_{\nu}}{{r^{ij}}^{2}}.
\end{align}
The $81$ components of the elasticity tensor $C_{\alpha\beta\mu\nu}$ can be reduced to $21$ independent values for physical media, and further to $3$ for physical media with cubic symmetry. Using Voigt notation we identify $C_{11} = \frac{1}{3}(C_{1111} + C_{2222} + C_{3333})$, $C_{12} = \frac{1}{3}(C_{1122}+C_{1133}+C_{2233})$, and $C_{44} = \frac{1}{3}(C_{2323}+C_{3131}+C_{1212})$. \dlp{} automatically reports the values of the elastic constants the 21 possible elastic constants when the required stress correlation terms are present. To compare with experimental data, from the elastic constants we can derive the bulk and shear modulus. The bulk modulus is
\begin{align}
    B = \frac{1}{3}(C_{11}+2C_{12}).
\end{align}
For the shear modulus there is a choice of averaging, $K^{V}$ (Voigt) and $K^R$ (Reuss) which give upper and lower bounds of the experimental results \cite{hill1952elastic}. These take the form
\begin{align}
    K^V &= \frac{C_{11}-C_{12}+3C_{44}}{5}, \\
    K^{R} &= \frac{5}{4/(C_{11}-C_{12})+3/C_{44}}.
\end{align}
Both $K$ and $B$ are reported for experimentally observed argon crystals. Note that the isothermal compressibility is often reported where $\chi = B^{-1}$, and the bulk modulus itself may also be more simply calculated from fluctuations in volume in an NPT ensemble as \cite{allentild}
\begin{align}
    B = \frac{\langle V \rangle k_{\mathrm{B}}T}{\langle V^2 \rangle - \langle V\rangle ^2}.
\end{align}

\subsubsection{Comparison to simulations and experiments}
\begin{figure}
    \centering
    \includegraphics[width=\columnwidth]{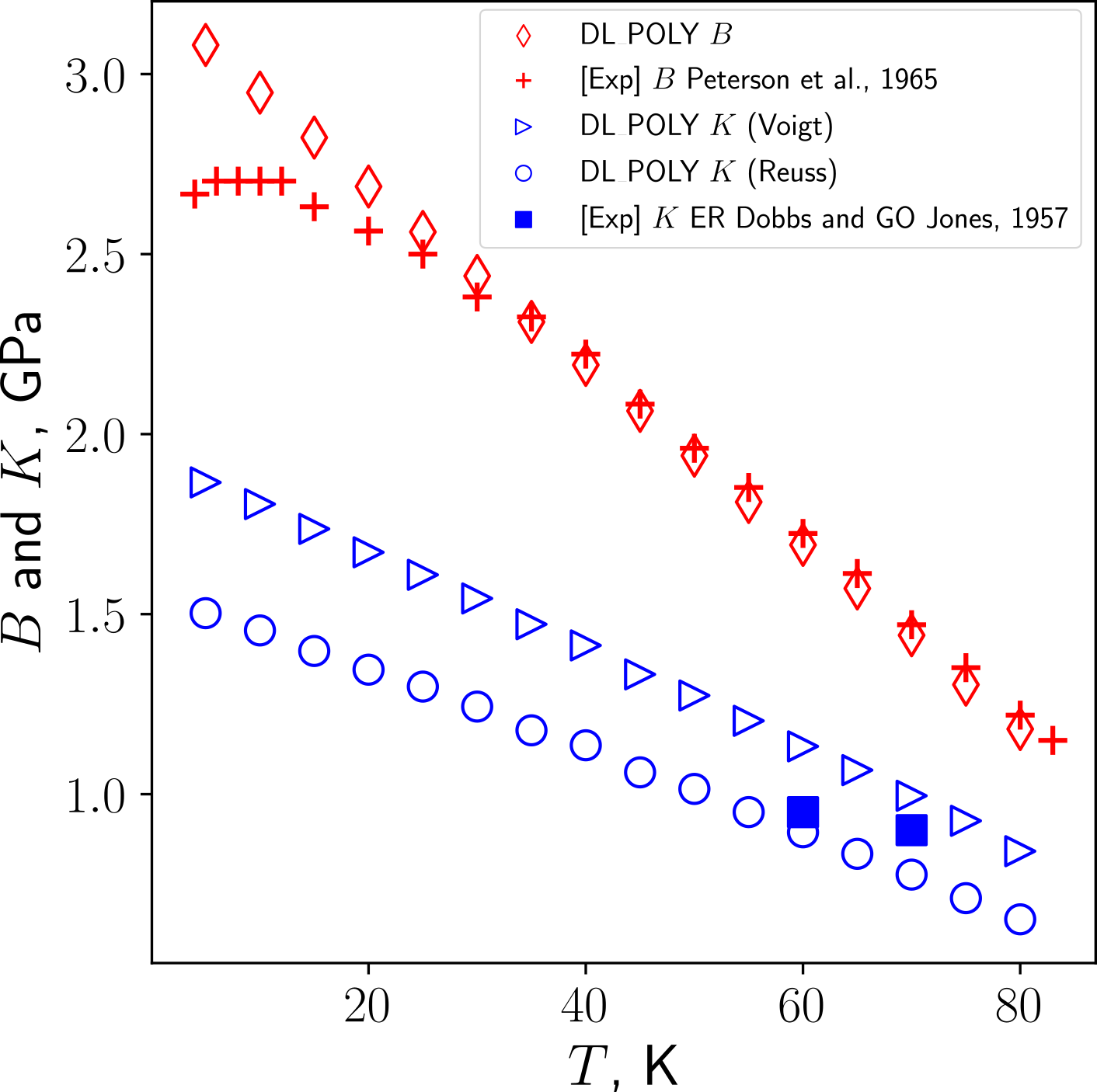}
    \caption{The bulk and shear modulus, derived from \dlp{} elastic constants (see Fig \ref{fig:sims-comp} for the raw data), compared with experimental data. Voigt and Reuss averaging give upper and lower bounds on the shear modulus respectively as expected. The Bulk modulus is also in good agreement Like \cite{pereverzev2022isothermal} we find a linear relationship in $C_{11}$ and $C_{12}$ with temperature and hence the same is true for $B$.} 
    \label{fig:exp-comp}
\end{figure}
 The following methodology was used in all our simulation results. An FCC Argon crystal was initialised and equilibrated in NPT, for $10^5$ steps (timestep $1$ fs) with a Nos\'{e}-hoover thermostat with thermo- and baro-stat couplings of $0.1$ ps and $0.1$ ps respectively, and a cutoff of $12$ \AA{}. The potential was the Lennard-Jones potential with the same parameters as \cite{clavier2017computation}. 
 
 After equilibration production runs (NVT) were simulated for $10^6$ steps unless otherwise specified. For these runs, the atoms were re-scaled to the average simulation cell from the equilibration phase. Statistical averages were calculated on a rolling window of $10$ ps updated every step, and correlations computed for a maximum lag time of $10$ ps, also updated every step using \dlp{}'s on-the-fly correlators. Everything else was kept the same. For each replicate $n=1,2,\ldots, 20$, this procedure was carried out with a different random seed for equilibration and production. The methodology is similar to \cite{clavier2017computation}, except they generate initial inputs through Monte-Carlo (MC) methods.

Fig. \ref{fig:sims-comp} compares \dlp{}'s calculated elastic constants with existing simulation results on argon FCC crystals, including MD and MC results, which are notably quite sparsely reported along the temperature axis. Previous simulation results are obtained from MC \cite{squire1969isothermal} and MD \cite{ quesnel1993elastic, pereverzev2022isothermal, clavier2017computation, thompson2022general}. We find that the values reported by \dlp{} are consistent with previous simulation results. The results from \citet{pereverzev2022isothermal} and \citet{squire1969isothermal} follow broadly the same trend in temperature, and the remaining results reported at $60$ K are also consistent with \dlp{}'s.

Given that we find consistent results for the individual elastic constants against simulated data we now compare with experimental values. For the bulk modulus ($B$) these follow the same averaging scheme. \dlp{}'s results, Fig \ref{fig:exp-comp}, bound the experimental values of \citet{dobbs1957theory} as expected. The bulk modulus is also in good agreement with \citet{peterson1966}. Although there is a notable plateau trend at $\lesssim 20$ K, which is not captured in simulations. Not that \citet{pereverzev2022isothermal} also finds a linear relationship at low temperature consistent with our results (see Fig \ref{fig:sims-comp}).

\subsection{Currents}
In \dlp{} 5's currents module is capable of calculating various collective properties resolved to user supplied ${\bf k}$-space vectors. To begin, the most basic functionality is the computation of ${\bf k}$-space density, as well as transverse and longitudinal currents. These are defined as \cite{balucani}
\begin{align}
    n({\bf k}, t) &= \sum_{i}e^{i{\bf k}\cdot {\bf r}^{i}(t)}, \label{k-density}\\
    {\bf j}_{L}({\bf k}, t) &= \sum_{i}({\bf v}^{i}(t)\cdot \hat{{\bf k}})\hat{\bf k}e^{i{\bf k}\cdot {\bf r}^{i}(t)}, \label{long-current}\\
    {\bf j}_{T}({\bf k}, t) &= \sum_{i}[{\bf v}^{i}(t)-({\bf v}^{i}(t)\cdot \hat{{\bf k}})\hat{\bf k}e^{i{\bf k}\cdot {\bf r}^{i}(t)}. \label{trans-current}
\end{align}

Equation \ref{k-density} may be used to determine the intermediate scattering function using the correlations
\begin{align}
    F({\bf k}, t) = \frac{1}{N}\langle n({\bf k}, t)n(-{\bf k}, 0)\rangle,
\end{align}
which can be used to determine thermal-conductivity in fluids \cite{cheng2020computing}, as well as the static structure factor as the $t=0$ correlation value. The dynamic structure factor may also be obtained using the Fourier transform of $F({\bf k}, t)$. Analogous correlations of ${\bf j}_L({\bf k}, t)$ and, ${\bf j}_T({\bf k}, t)$ may be used to determine transverse and longitudinal propagating modes. These collective properties relate to the overall motion of the particles of the system. Collective excitations form an integral and well known part of the theory of solid and gaseous states, but for liquids, which combine strong interactions with dynamical disorder, this has historically not been the case. 

Alongside these currents, \dlp{} can also calculate the ${\bf k}$-space energy density, energy currents, and ${\bf k}$-dependent stress tensor. 

The energy density is given by
\begin{align}
    e({\bf k}, t) = \frac{1}{2}\sum_{i}E^{i}e^{i{\bf k}\cdot {\bf r}^{i}(t)}.
\end{align}
and the ${\bf k}$-dependent stress and energy currents from equations \ref{k-stress} and \ref{eng-current}.
\begin{align}
    {\bf \sigma}_{\alpha, \beta}({\bf k}) &= \sum_{i}\biggl( mv^{i}_{\alpha}v^{i}_{\beta}-\frac{1}{2}\sum_{j\neq i}\frac{r^{ij}_{\alpha}r^{ij}_{\beta}}{|r^{ij}|^2}P({\bf k}, {\bf r}^{ij}) \biggr)e^{i{\bf k}\cdot {\bf r}^{i}(t)}, \label{k-stress}\\
    {\bf q}_{a}({\bf k}) &= \frac{1}{2} \sum_{i} \biggl[ E^i {\bf v}^{i}_{a} - \frac{1}{2}\sum_{j\neq i}\sum_{b}({\bf v}^{i}_{b}+{\bf v}^{j}_{b})({\bf r}^{ij}_{a}{\bf r}^{ij}_{b}/|{\bf r}^{ij}|^2)P({\bf k}, {\bf r}^{ij})\biggr]e^{i{\bf k}\cdot {\bf r}^{i}(t)}, \label{eng-current}\\
    \text{where } P_{k}({\bf k}, {\bf r}) &= |{\bf r}|\frac{\partial U(|{\bf r}|)}{\partial r}\frac{1-e^{-i{\bf k}\cdot {\bf r}}}{i{\bf k}\cdot {\bf r}}
\end{align}
A user may request correlations of combinations of these various currents (component wise). Each ${\bf k}$-point specified by the user will result in one correlation matching these requests in the \texttt{CONTROL} file. One use case for current correlations is calculating phonon dispersion curves. These dispersions can be calculated in \dlp{} via the calculation of current correlations
\begin{align}
    C_{L}({\bf k}, t) &= \langle {\bf j}^{z}_{L}({\bf k}, t) {\bf j}^{z}_{L}({\bf k}, 0)\rangle, \\
    2C_{T}({\bf k}, t) &= \langle {\bf j}^{x}_{T}({\bf k}, t) {\bf j}_{T}^{x}({\bf k}, 0)\rangle\nonumber\\&+\langle {\bf j}^{y}_{T}({\bf k}, t) {\bf j}_{T}^{y}({\bf k}, 0)\rangle.
\end{align}

From these correlation functions, the spectra are found from the Fourier transforms $C_{L}({\bf k}, \omega)$ and $C_{T}({\bf k}, \omega)$. The frequencies corresponding to maxima of these Fourier transforms of each wave-vector are interpreted as the spectrum of collective modes \cite{balucani,yangprl}. 

In liquids in particular, it is possible to observe an interesting feature of spectra not seen in solids: $k$-gap in the solid-like transverse spectrum meaning a threshold value below which no such propagating modes exist \cite{gmsreview,yangprl}. The $k$-gap can be observed by plotting the relationship between these maxima and $k=|{\bf k}|$. 

\begin{figure}
    \centering
    \includegraphics[width=\columnwidth]{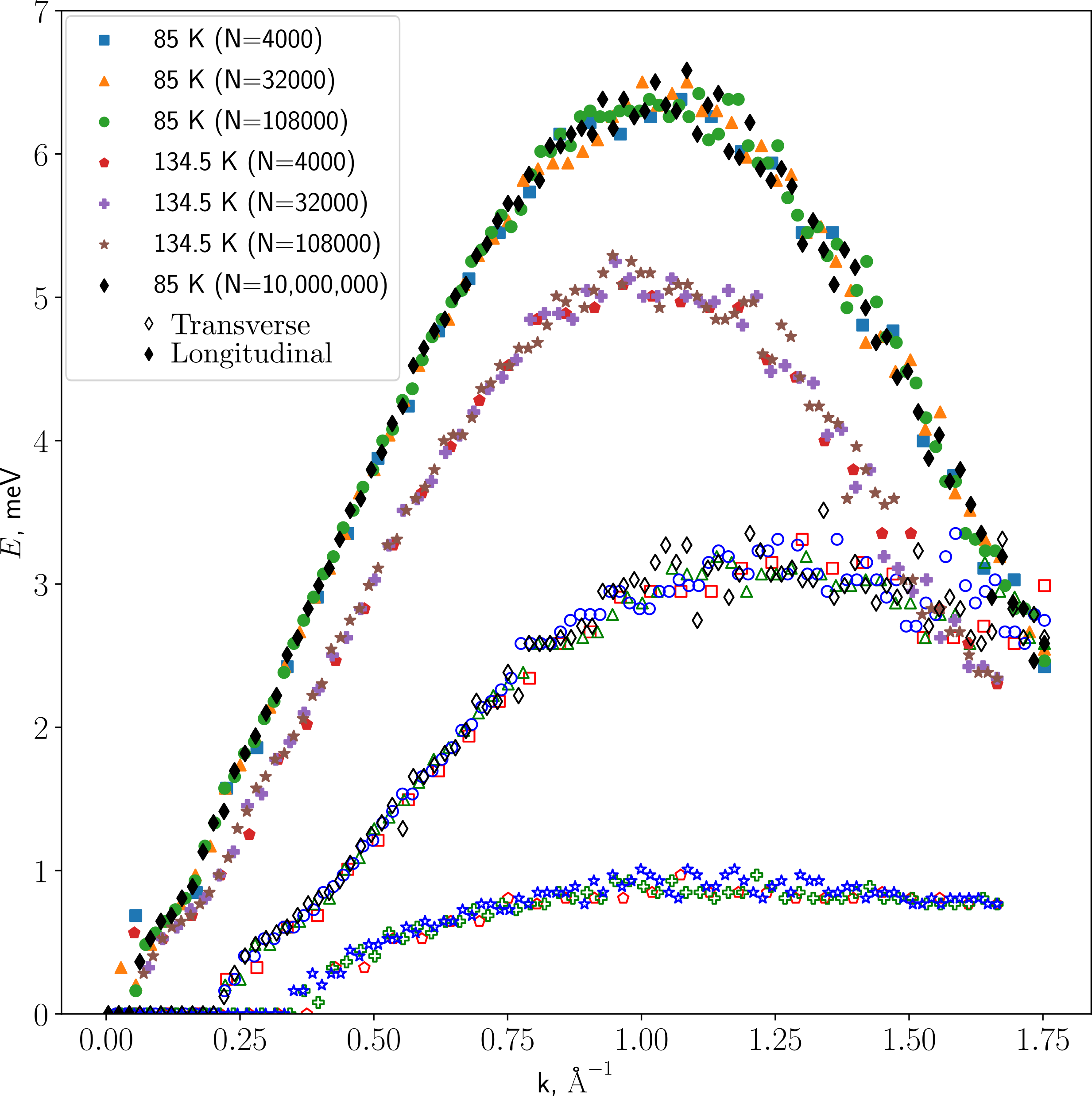}
    \caption{Dispersion curves for Argon at 85 K and 134.5 K for system sizes 4,000, 32,000, 108,000, and 10,976,000.}
    \label{fig:k-gap}
\end{figure}

To calculate these data in \dlp{} we first equilibrate an argon system of 4,000 atoms in NPT to the desired temperature. We then expand to other desired system sizes using \dlp{}'s \texttt{nfold} capability. We then calculate the correlations functions on-the-fly over $10^{6}$ simulation steps ($1$ fs per step), for all physically reasonable ${\bf k}$-points, $k{\bf z}$. Our correlations are resolved for $10240$ points per block, at a correlation frequency of every $10$ steps (lags up to $102400$ fs). The resulting correlation functions are first averaged across $20$ independent trajectories. The frequency maxima are then found by taking Fourier transforms of these averaged correlations. The result of this is shown in figure \ref{fig:k-gap} where we see the increasing $k$-gap for the transverse mode for temperatures of $85$ K and $134.5$ K. 

\subsection{Rigid body correlations}
In \dlp{} rigid bodies can be defined as part of user input in the \texttt{FIELD} file. Molecules may have single or multiple rigid body components. During a simulation each rigid body has position, velocity and angular velocity data which may be correlated in the same manner as per-atom correlations. That is \dlp{} will determine distinct rigid body types based on distinct rigid body components within distinct molecular types, calculate requested correlations at runtime, and average these results for each. 

\begin{figure}
    \centering
    \includegraphics[width=\columnwidth]{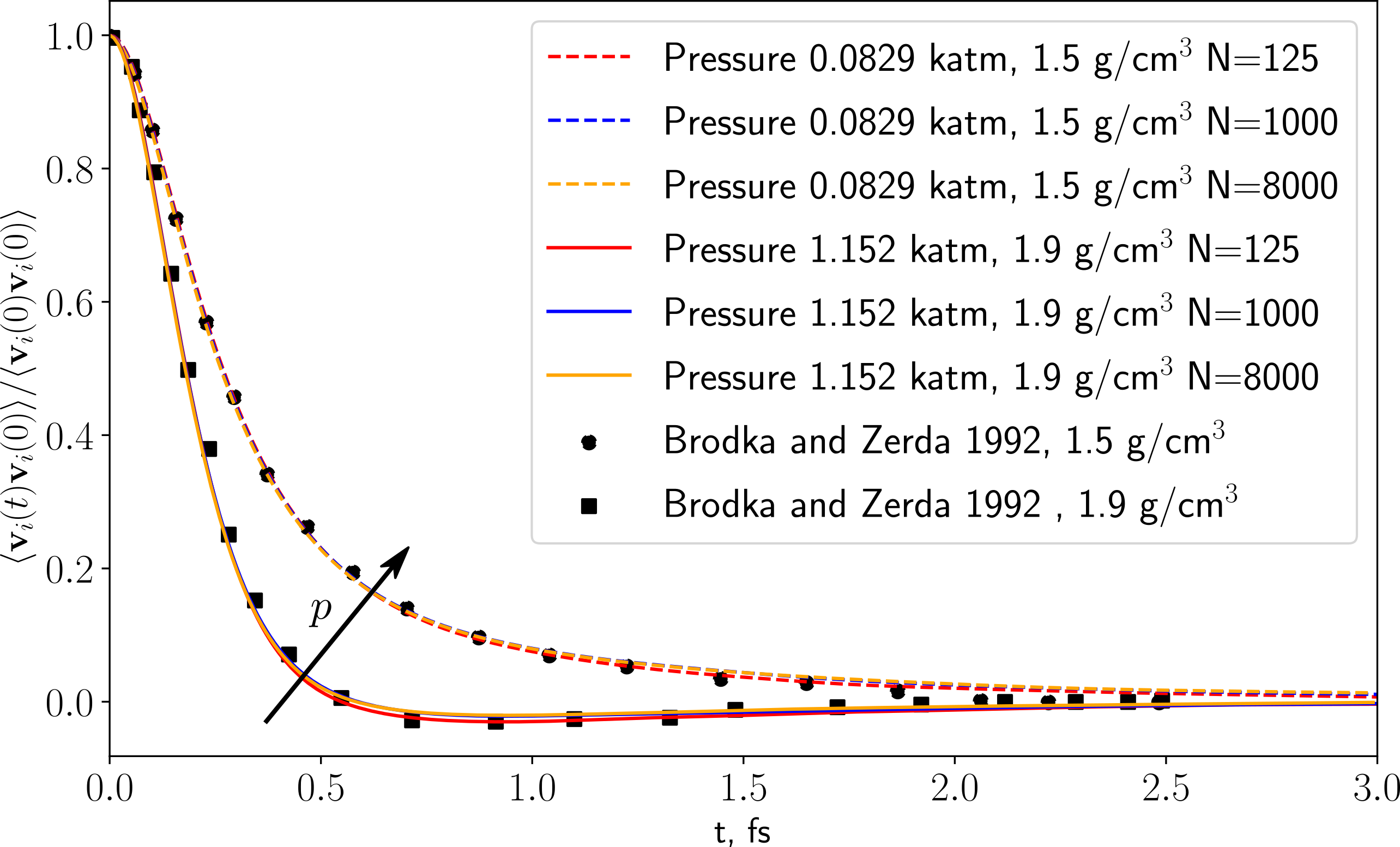}
    \caption{Comparison to Brodka and Zerda's results (digitised) for SF${}_{6}$ rigid body VAFs \cite{brodka1992molecular}. Temperature 296 K with pressure and density indicated in the legend. $N$ indicates the rigid body count.}
    \label{fig:sf6}
\end{figure}

Using this on-the-fly correlation functionality we are able to reproduce the results of \citet{brodka1992molecular} for SF${}_{6}$. We do this by simulating rigid bodies of SF${}_{6}$ molecules with the same Lennard-Jones potential parameters. For equilibration we set the temperature to 296 K, and pressure to the experimental values listed for the pressure by Brodka and Zerda. We simulate in NPT for $100$ ps to ensure equilibration. The resulting densities were similar to the $1.5$ and $1.9$ g/cm${}^{3}$ listed in their results. We then rescaled our resulting configurations to match those densities exactly, and then proceed in NVE collecting statistics over $1$ ns simulation steps. Brodka and Zerda use $1000$ equilibration steps and $4200$ production steps at a higher time-step ($5$ fs vs. $1$ fs). Our results for the VAF are shown in Figure \ref{fig:sf6} which match well. 

\begin{figure}
    \centering
    \includegraphics[width=\columnwidth]{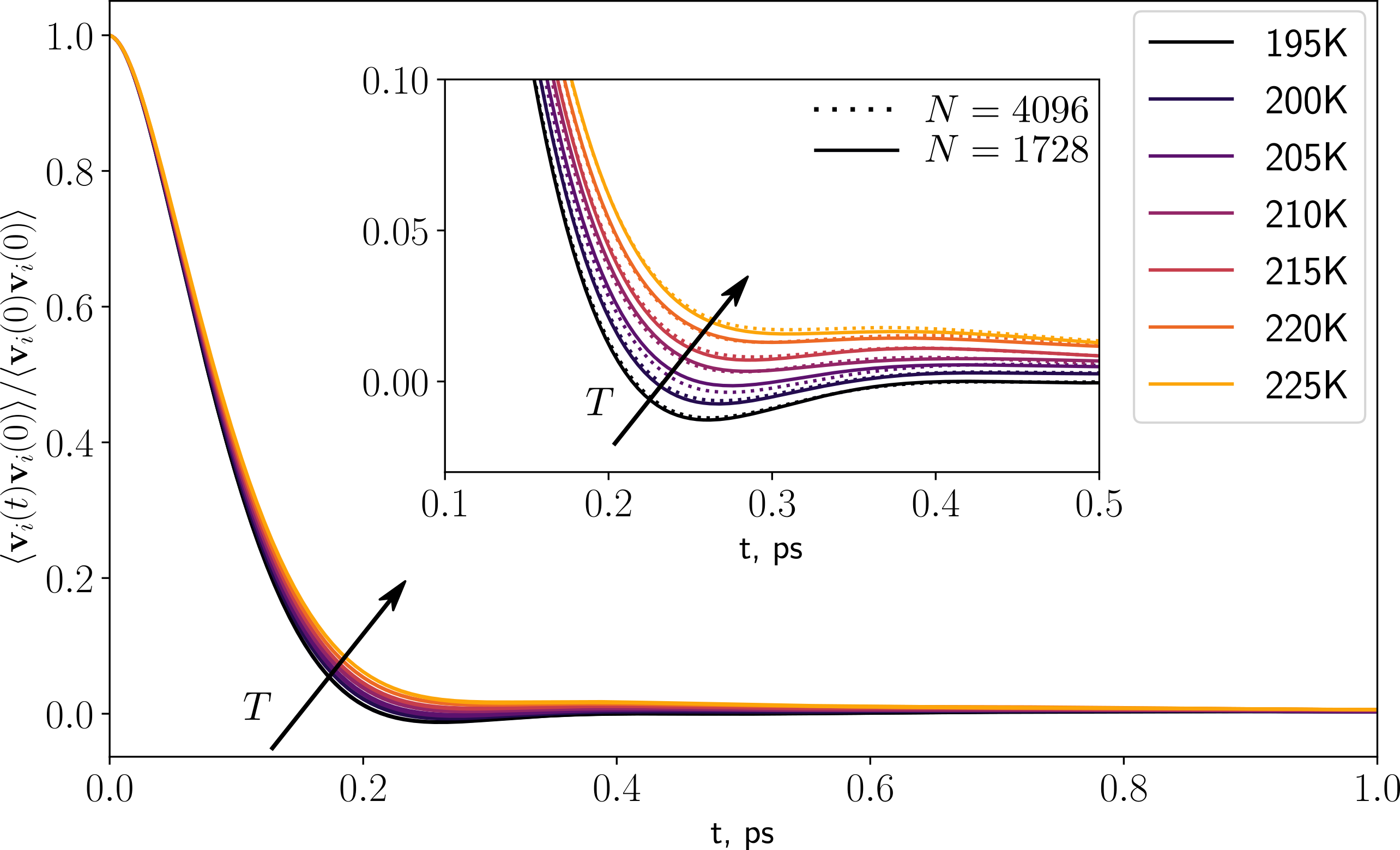}
    \caption{VAF for CH${}_{4}$ rigid molecules at $900$ bar for increasing $T$, showing the crossover at the Frenkel line.}
    \label{fig:ch4}
\end{figure}

The VAF can also be used as a definition of the Frenkel line in supercritical fluids \cite{Cockrell2021b,brazhkin2013}, which marks the crossover between liquid-like and gas-like particle dynamics and thermodynamics. Using the rigid body VAF functionality we perform this analysis for rigid body CH${}_{4}$ molecules. We follow the process of \citet{Yang2015}, using the same potential \cite{skarmoutsos2005investigation}. We first equilibrate our system in NPT at the desired $P$ ($900$ bar) and $T$ conditions for $50$ ps and then collect statistics over a further $50$ ps. As before we find the equilibrated simulation densities to be approximately the same as the NIST experimental data for supercritical CH${}_{4}$ \cite{nist}, and for increasing $T$ there is a loss of the VAF minimum as shown in figure \ref{fig:ch4}.

\subsection{Benchmarks and Trade-offs}
We have identified a variety of applications of the on-the-fly correlator in \dlp. In each case the calculated correlations forgo the need for storing the underlying trajectory. However for different use-cases this trajectory is more or less significant. Consider for example the viscosity and thermal conductivity. \dlp, analogously to many other packages, already calculates the heat flux and stress tensor at each time-step. This data (the 3 components of heat flux or 6 components of the stress tensor per step) is small and does not scale with the system size. Therefore, the printing of the full trajectory (velocities and forces) can be avoided in these cases. For the VAF the velocities are required, while for currents the reduced data are still much more significant. In these cases the written data scales with the system size explicitly (atom velocities) or implicitly (possible ${\bf k}$-points). For the thermal-conductivity and viscosity, the use of the correlator is a significant convenience (and a well defined reproducible workflow) but not a large performance boost. We can realise the full benefit of avoiding trajectory storage and benchmark our work where heavy I/O can be avoided, for example computing the VAF.
\subsubsection{Strong and weak scaling: the velocity autocorrelation function}

\begin{figure}
    \centering
    \includegraphics[width=\columnwidth]{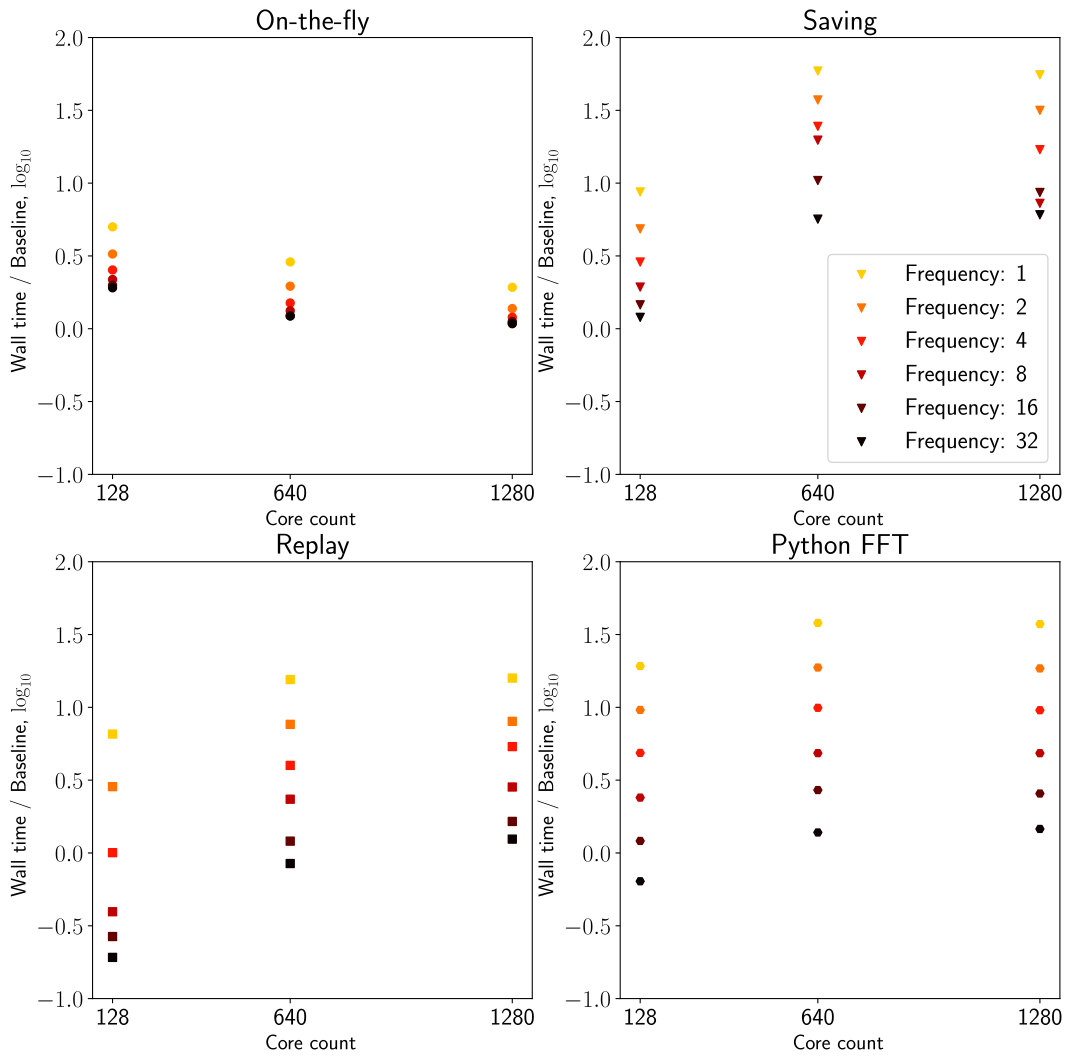}
    \caption{Strong scaling benchmarks (calculating the VAF) with 1,000,000 LiF atom simulations. Frequency indicates the sampling interval i.e. either correlation frequency or trajectory writing frequency. In all cases the y axis is the wall time relative to the baseline simulation (without trajectory writing or correlation) on a log scale. The baseline wall times were 123.19 s, 66.14 s, and 75.4 s for the 128, 640, and 1280 core simulations. The absolute wall times for the simulations can be found in Table \ref{strong-sim} and the absolute wall times for the post-processing methods in Table \ref{strong-post}.}
    \label{fig:vaf_strong}
\end{figure}

\begin{figure}
    \centering
    \includegraphics[width=\columnwidth]{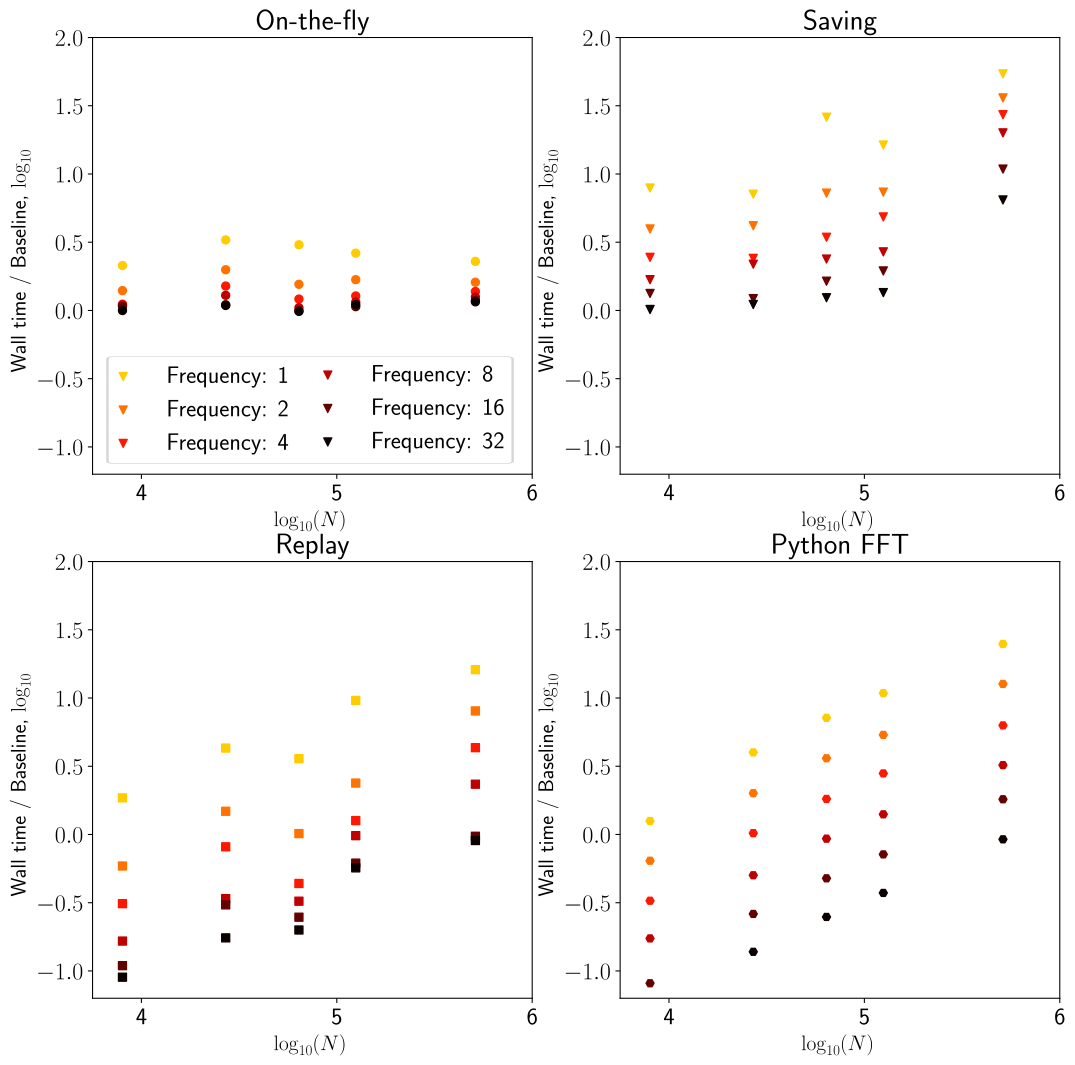}
    \caption{Weak scaling benchmarks (calculating the VAF) with 1,000 LiF atoms per core. Frequency indicates the sampling interval i.e. either correlation frequency or trajectory writing frequency. In all cases the y axis is the wall time relative to the baseline simulation (without trajectory writing or correlation) on a log scale. The baseline wall times were 16.0 s, 16.43 s, 20.34 s, 26.86 s, and 49.89 s for the $N=8000$, $27000$, $64000$, $125000$, and $512000$ atom simulations. The absolute wall time data can be found in Table \ref{weak-otf} for the on-the-fly simulations, in Table \ref{weak-save} for trajectory saving, in Table \ref{weak-replay} for the trajectory replay post-processing method, and in Table \ref{weak-python} for the Python FFT based post-processing.}
    \label{fig:vaf_weak}
\end{figure}

\begin{figure}
    \centering
    \includegraphics[width=\columnwidth]{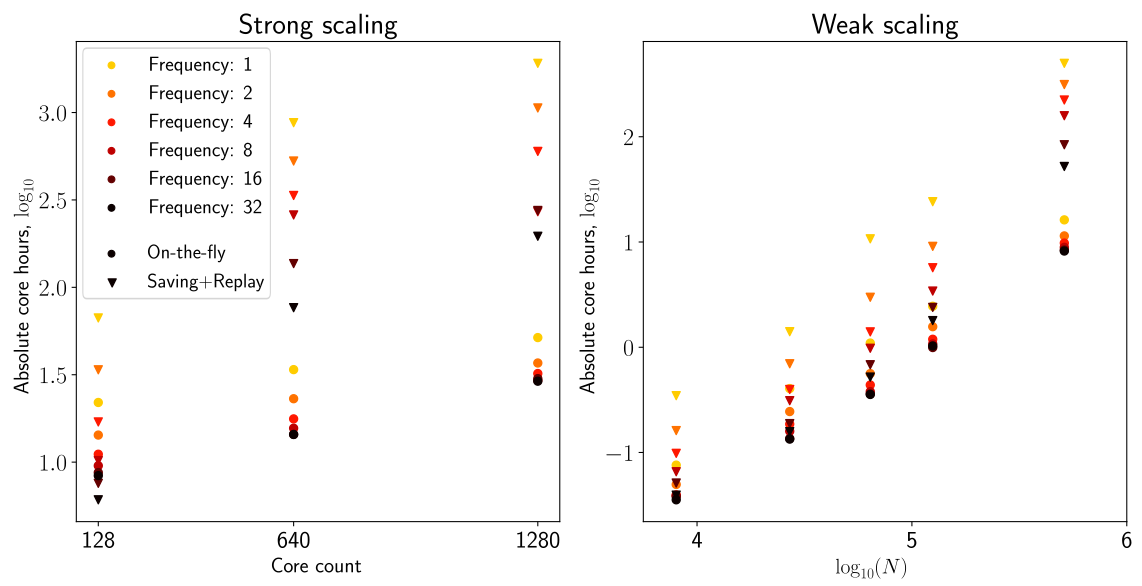}
    \caption{Core hour analysis for the strong and weak benchmark data presented in figures \ref{fig:vaf_strong} and \ref{fig:vaf_weak}. The circular scatter points indicate the full on-the-fly workflow core hour cost. The downward triangles indicate the combination of a simulation with trajectory saving followed by calculating correlations by trajectory replay (which is generally the fastest post-processing method presented here).}
    \label{fig:vaf_core_hours}
\end{figure}

In figures \ref{fig:vaf_strong} and \ref{fig:vaf_weak} we show strong and weak scaling benchmarks for calculating the VAF of molten LiF with $10^{6}$ atoms, using the Buckingham potential in the NVE ensemble. That data for all scatter points can be found in tabulated form in Appendix \ref{benchmark-data}. Additionally we present that same data in figure \ref{fig:vaf_core_hours} in the format of CPU core hours for the full workflows (on-the-fly versus saving the trajectory and analysing the trajectory).

In all cases the simulations were identical except for the mutually exclusive modes of trajectory writing or correlating. The simulations were run for 1000 time-steps with a cutoff of 8 \AA{} and an SPME precision of $10^{-6}$. STATIS and OUTPUT were suppressed by setting their output frequency greater than the number of steps. In the figure captions ``Frequency $f$'' refers to a sampling interval of $f$ simulation time steps i.e. the frequency of trajectory saving, or correlation. Post-processing of the trajectories was performed by enabling \dlp's replay feature (also using the on-the-fly correlator) and in Python using FFTW \cite{FFTW} (also using simple multiprocessing across atoms) to cover basic user post-processing analysis strategies. All simulations were performed on the Sulis Tier 2 HPC system utilising the standard ``compute nodes''. These were Dell PowerEdge R6525 compute nodes each with 2 x AMD EPYC 7742 (Rome) 2.25 GHz 64-core processors (128 cores per node) with 512 GB DDR4-3200 RAM per node. The interconnect was Mellanox ConnectX-6 HDR100 (100 Gbit/s) InfiniBand and the storage system utilised was General Parallel File System (GPFS)
. The default OpenMPI IO subsystem was used for these benchmarks. Sulis also offers ROMIO \cite{romio} which had little impact. In the case of the trajectory saving and replay benchmarks care was taken to select the best performing split of MPI processes as readers and writers. These are subsets of the same MPI count used for domain decomposition. They can work independently on the same I/O file. All MPI tasks depend on the I/O dedicated MPI tasks and no compute can take place until all I/O has been completed as commitment to the memory of the I/O subsystem. This can be controlled in \dlp's CONTROL file directly. Typically this meant using equal or less than the node size (128).

The strong scaling benchmarks show that \dlp's correlator module is able to make use of the increasing core counts well. Especially at higher frequencies when compared to writing the trajectory, as expected given \dlp's domain decomposition. In comparison purely saving the trajectory incurs a much larger performance impact (when saving every step the data measures $\sim 285$ GB). Except for the cases of low frequency (64-128 steps), and more widely for the single node case. As remarked above, but not shown in these data, the benchmarks for trajectory saving and replay depend heavily on the user's selection of reader and writer processes. A clear benefit of the on-the-fly method is this is simply not a concern. For example in the worst case of the 1280 core simulations and saving every step, choosing 64/1280 cores as writers resulted in the reported wall time, 3019 s. But choosing 1280/1280 cores as writer results in a wall time of 5513.36 s. Finally note for a complete picture of the workflow the wall times for saving the trajectory must be added to one of the post-processing modes. Consider the best case for saving the trajectory (128 cores and 128 saving frequency). The simulation saving the trajectory took 154 s and the post-processing wall times were 64 s or 32 s for replay and the FFT method respectively. The on-the-fly wall time was 202 s. A $\sim 5\%$ saving or $\sim 8 \%$ loss in each case. 

Both post-processing methodologies are dominated by reading the trajectory data rather than the computation of the correlation. This can be seen by the poor scaling with increased core count which we include for complete picture. Further improvements could likely be obtained by using dedicated data-processing partitions where available.

The weak scaling benchmarks show a similar picture. A tight spread between the frequencies using the on-the-fly method and a widening gap between the heavy I/O and light I/O when saving or post-processing the data. Again we also see the impact of multi-node I/O where the wall times are impacted depending on the user's choice of read and writer processes.

Finally, the core hour comparison in figure \ref{fig:vaf_core_hours} again indicates the benefits that can be obtained using the on-the-fly method. In this case we see the impact of the combined cost of trajectory saving and analysis compared with using the on-the-fly method. The increased core hours come from I/O operations. Depending on accuracy requirements use of compression schemes such as ZFP could reduce this by 30\% (64-bit, lossless) or more than 90\% (32-bit, lossy). In the worst cases (strong scaling, 1280 cores) the over head is approximately between $10^2-10^{1.5} \approx 68$ core hours and $10^{3.5}-10^{1.5}\approx 3130$ core hours. At 30\% and 90\% data reductions this could lead to overheads of 48 core hours and 312 core hours respectively.

\subsubsection{When to use on-the-fly correlations}
\dlp's on-the-fly correlations module provides the ability to calculate correlations without saving trajectories. We have demonstrated that this can be a significant saving, particular in calculating the VAF. In general on-the-fly correlations make efficient use of \dlp's parallelisation, with low overhead. The primary use case is for large systems where analysis would normally require the trajectory to be written. Especially where analysis would dictate frequent writing. Concrete examples do depend on the hardware used. In general it is notable that when saving a significant trajectory a user must think carefully about the I/O setup of the job (read and writer splits).

However by bypassing storage of the trajectory the user is left unable to conduct further analysis without re-running the simulation. This could manifest from the need for unforeseen analysis or incorrect parametrisation. On this latter point an example would be computing a correlation to a far too short time scale, cutting off important features for analysis, or too infrequent correlation leading to poor resolution. One solution is to conduct preliminary analysis on small systems where I/O is negligible in order to determine the analysis requirements relating to correlation parameters. In this case the on-the-fly correlator can then be used for much larger systems. Or paying the price of trajectory storage for a smaller sample of large simulations, and using the on-the-fly correlator to about additional replicates for good statistics. A middle ground can be found in many cases. That is by saving the trajectory coarse enough that the runtime impact is moderate in order to preserve data for later use, whilst calculating certain properties on-the-fly at a much finer resolution. It is then down to the user to determine a reasonable saving frequency for unanticipated analysis, which is always a problem.

An advantageous alternative for the \dlp{} user results from \dlp's replay feature. That is a HISTORY file can be fed back into \dlp{} to re-calculate statistics, including new ones at finer or coarser scales, leveraging the domain decomposition and parallel I/O in \dlp. As shown in the benchmarks this has a competitive performance compared to using fast Fourier transform methods accessible with simple scientific programming as core counts and/or problem sizes increase. Of course there is a trade-off as well. Simple highly specific targeted analysis can obtain results with little overhead. Replaying a HISTORY file however, does incur overhead as \dlp is doing more than required.

In general across MD software on-the-fly methods will be subject to these concerns. But this will also change depending on the parallelisation strategy. \dlp{} is a CPU based software. When heterogeneous compute systems can be leveraged (such as CPU-GPU systems) trajectory saving could become less cumbersome if the CPU can largely write out data whilst the GPU does the bulk of the MD simulation workload. However the efficiency of this setup will depend on the relative speed of the CPU write-out (plus data transfer penalties) to the GPU compute speed. In cases where this is a bottleneck (e.g. high time-step resolution analysis), on-the-fly methods could be utilised on the GPU side to reduce the communication of data. Another way to avoiding I/O/transfer overheads.

\section{\dlp{}, massive parallelism and scalability}

The \dlp{} project was conceived in 1993 by W Smith and originally released in 1994 as \dlp\_2 \cite{smith1996dl_poly_2, smith2002dl_poly} (now known as \dlp{}\_CLASSIC \cite{dlpoly-classic-code}) using a replicated data (RD) approach as its underlying parallelisation strategy \cite{smith1994parallelI, smith1994parallelII}, written in Fortran77 \cite{brainerd1978fortran} and using MPI \cite{walker1996mpi}. Later in 2003, led by I. Todorov, with core work carried out by W. Smith and I. Bush, \dlp{} 3 was released which developed the foundation of the domain decomposition (DD) parallelisation strategy \cite{todorov2004dl_poly_3, todorov2006dl_poly_3}, including a Smooth Particle Mesh Ewald (SPME) method suitable for the \dlp{} DD \cite{bush2006daft}. This strategy provided for an extremely efficient memory distribution across a variety of test systems \cite{todorov2006dl_poly_3}.

\begin{figure}
    \centering
   \includegraphics[width=\columnwidth]{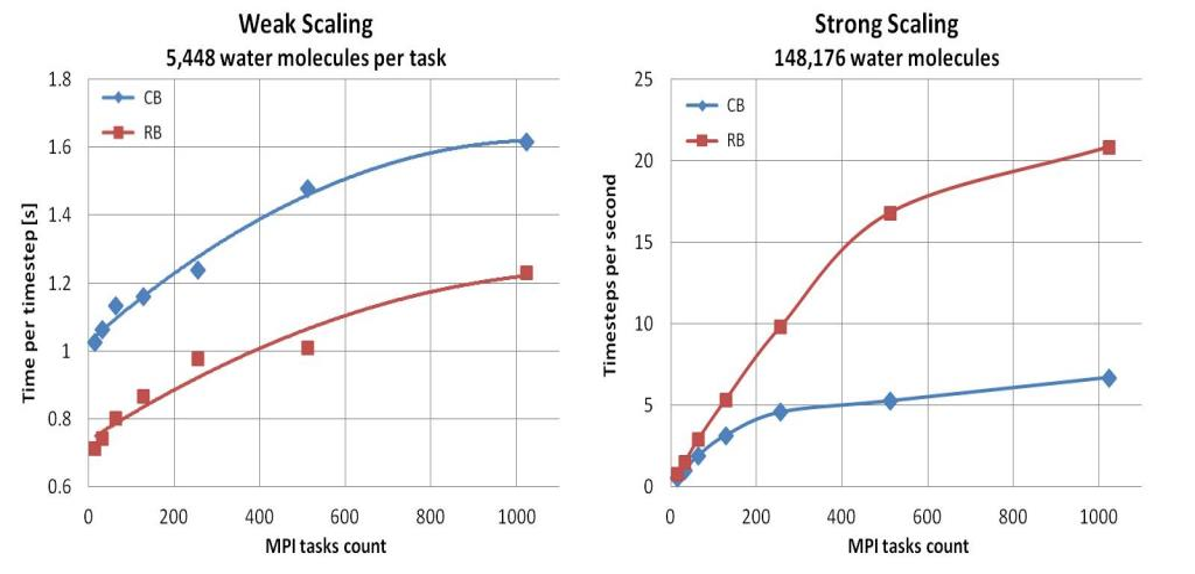}
    \caption{Weak and strong scaling benchmarks of SPC water systems.  This data was first produced with \dlp{} 4.04 in May 2013 on UK's National HPC service HECToR (Cray XE6).}
    \label{water}
\end{figure}

The release of \dlp{}\_4 included a major rewrite providing an extension of particle dynamics to rigid body dynamics \cite{todorov2013}, inclusion of two-temperature thermostat functionality \cite{seaton2021} for high-energy events and the inclusion of parallel I/O \cite{cug2010}. The rigid body dynamics provided a natural extension of the parallel performance capabilities for biological and biochemical systems in which constrained bond dynamics was a bottleneck. The enhancement in performance of rigid body (RB) dynamics and constrained bond (CB) dynamics is demonstrated in Fig. \ref{water} where both weak and strong scaling tests were performed on the same SPC water system and model with scaled sizes in the case of weak scaling. Further developments in 2012 led to the inclusion of Dissipative Particle Dynamics integration \cite{shardlow} and a three seeded random number generation \cite{seaton2013} for using repeatable random series for stochastically dependent thermostating, kinetic fields and pressure tensors in a fully deterministic manner across domains and time series.

A few years ago an extensive refactoring effort took place that is the base of the current version of \dlp{}. These changes were introduced mainly due to the need for being able to add new complex tasks with maximum code reuse and allow various easy optimisations. The main changes introduced were: i) using object oriented Fortran 2008 as the minimal standard required, ii) restructuring the code to be fully modular, with all functions and subroutines exchanging data only via parameters rather than via modules - the removal of implicit \texttt{common} blocks. Only one singleton remains for dealing with I/O for errors. iii) introducing an automated test system with both regression and unit tests. Testing was extended from 24 tests to over 200 tests. iv) atomic configurations are stored in an array of structures that is cache friendly with respect to positions and forces (although this has a non-trivial trade-off with vectorisability of the code), v) electrostatics was rewritten to allow per-particle decomposition and allowing the computation of long range contributions from any power potential, see next section for details. vi) The \texttt{CONTROL} file format was redesigned (see appendix \ref{new-control}). In general all entries now are consistently of the form \texttt{token-value-unit} (where applicable), STATIS, RDF and other output files can be written in the yaml format \cite{ben2009yaml} for easier postprocessing, vii) a modern software infrastructure is in place with full CI/CD via gitlab.com, vi) license was changed to an open-source license GPL-3.0, viii) dlpoly-py was developed \cite{dlpoly-py} - a companion python package that can read inputs and outputs of \dlp~and initiate simulations. Without this extensive refactoring, implementation of the Empirical Valence Bond \cite{scivetti2020} and the new Chemshell interface \cite{lu2018} would not have been possible. 

\subsection{SPME refactoring}

The Ewald method \citep{Ewald1921} is a method for the calculation of long-range (conditionally convergent until infinity) forces. This can be achieved in a number of ways, the usual physical description being the introduction of a screening charge distribution and a correction to account for the screening so as to accelerate the summation of the long-range component in Fourier space. This can also be seen as performing a sum from $0\rightarrow{}\alpha{}$ in real-space and from $0\rightarrow{}\alpha'{}$ in reciprocal space to construct a full sum of $0\rightarrow{}\infty{}$ in finite time ($\lim\limits_{x\rightarrow{}0}\frac{1}{x}=\infty{}$). 

The smooth particle mesh Ewald (SPME) approach by \citet{essmann1995smooth} uses cardinal B-Splines to interpolate the complex exponentials that result from the above approach in a regular fashion. This allows the use of a standard Fast Fourier Transform (FFT), so reducing the scaling of the method with system size to $O(N\log(N))$. However this is at the cost of some accuracy due to the interpolation, though this can be negated by using higher order splines to provide a better description of the exponentials.

The SPME approach in \dlp{} has been refactored to enable computation of per-particle contributions to energies, forces and stresses. The calculation of per-particle contributions has been used to implement the computation of Green-Kubo thermal conductivity with Ewald methods (See section \ref{therm-con}).

This new method coincidentally enables the expansion to not only alternative orders of $\tfrac{1}{r^n}$ potentials, but also derivatives of arbitrary order, thereby paving the way to correct multipolar expansions of arbitrary order. Direct computations of higher-order potentials allow ``exact'' computation of, e.g. Lennard-Jones (6-12) \cite{LJOrig} potentials without cut-offs, enabling calculation of the effect of these cutoffs on actual dynamics. Currently implemented are functional forms for: $r^{-1}$, $r^{-2}$, $r^{-6}$, $r^{-12}$ and a general form for $r^{-n}$ and their derivatives for both the real- and reciprocal-space parts (see appendix \ref{spme-kernel}).

\begin{figure}
    \centering
   \includegraphics[width=\columnwidth]{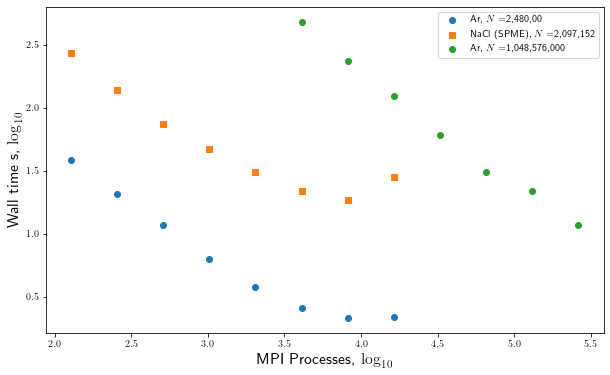}
    \caption{Strong scaling benchmarks for \dlp{} for Argon (2-body) and NaCl (2-body and SPME electrostatics) systems. Benchmarks were conducted on Archer2 \cite{archer2} all are timed over 1ps simulation time. The system sizes were of the order of millions and billions for Argon, and millions for NaCl. Processor counts vary from 128 in the smaller systems, up to 262,144 for the 1 billion atom system. Our largest benchmark represents a 35\% concurrent utilisation of Archer2 \cite{archer-hardware}}
    \label{strong1}
\end{figure}

\subsubsection{Benchmarks of the Refactored Code}
As part of the \dlp\_5 developments, we have recently benchmarked 2-body and SPME-enabled systems composed of $\sim\!\!2$ million atoms and 2-body simulations composed of $\sim 1$ billion atoms across increasing MPI process counts on Archer2 \cite{archer2}. We examine processor counts ranging from 128 in the small systems to 262,144 in the larger system. Representing a 35\% concurrent utilisation of Archer2's 750,080 CPU cores \cite{archer-hardware}. The results can be found in Fig. \ref{strong1}. Primarily the results continue to demonstrate the scalability of \dlp. But secondly, it also demonstrated the scalability of the newly refactored MD algorithms, with the reworked DD memory distribution that favoured purely local communications and minimised all-to-all communications as much as possible. A comprehensive survey of \dlp{} performance on modern hardware can be found in \cite{guest2021dl_poly}.

\section{Conclusions}
We have used the paradigm of on-the-fly calculations to develop a general purpose correlation module in \dlp{} 5. This functionality exposes key properties derived from simulation data which may be composed into correlation functions by the user. This method provides a convenient way to perform correlation analysis as well as eliminating the cumbersome I/O operations inherent in saving trajectory data for post-processing. It also presents the continuous availability of correlation function during simulation time. This saving becomes significant for high temporal resolution correlations defined at the atomic or molecular level such as the VAF.

\dlp{} continues to show the same scalability for large CPU counts as past versions. Our correlation implementation naturally takes advantage of the DD algorithm and multi-processing architecture used in \dlp{} to distribute the calculation of atomistic correlations. We aim to continue the development of on-the-fly methods in and beyond \dlp{} 5. For example, to date we have exposed rigid body molecular positions, velocities, and angular velocities. A natural extension is to correlate properties of flexible molecular species and other atomic groupings more generally. In particular the more complex sub-components of flexible molecules such as atomic bonds or angle based interactions etc.

\section*{Author contributions}
% NB https://www.elsevier.com/researcher/author/policies-and-guidelines/credit-author-statement
\textbf{H. L. Devereux:} Methodology; Software; Validation; Formal analysis; Investigation; Data Curation; Writing - Original Draft; Writing - Review \& Editing; Visualization.
\textbf{C. Cockrell: } Methodology; Software; Validation; Formal Analysis; Investigation; Writing - Original Draft; Writing - Review \& Editing.
\textbf{A. M. Elena: } Conceptualization; Methodology; Software; Validation; Formal Analysis; Investigation; Resources; Data Curation; Writing - Original Draft; Writing - Review \& Editing; Supervision; Project administration; Funding Acquisition.
\textbf{Ian Bush: } Software; Validation; Data Curation; Writing - Review \& Editing;
\textbf{Aidan B. G. Chalk: } Software; Validation; Data Curation.
\textbf{Jim Madge: } Software; Validation; Data Curation.
\textbf{Ivan Scivetti: } Software; Validation; Data Curation.
\textbf{J. S. Wilkins: } Software; Validation; Data Curation; Writing - Original Draft; Writing - Review \& Editing.
\textbf{I. T. Todorov: } Conceptualization; Methodology; Software; Validation; Formal Analysis; Investigation; Resources; Data Curation; Writing - Original Draft; Writing - Review \& Editing; Supervision; Project administration; Funding Acquisition.
\textbf{W. Smith: } Software; Conceptualization; Methodology; Validation; Data Curation.
\textbf{K. Trachenko: } Conceptualization; Methodology; Validation; Formal analysis; Investigation; Resources; Writing; Writing-Review \& Editing; Supervision; Project administration; Funding acquisition.

\section*{Data statement}
The \dlp{} 5 source code is hosted on GitLab \cite{dlpoly-code} (GPLv3.0). The input files and submission scripts to reproduce the statistical data for our on-the-fly methods can be found on Gitlab \url{https://gitlab.com/apw951/dl_poly_workflows/-/tree/dlpoly_5_preprint?ref_type=tags}.
\section*{Acknowledgements}
We are grateful to EPSRC (for grant No. EP/W029006/1). This research utilised Queen Mary's Apocrita HPC facility, supported by QMUL Research-IT http://doi.org/10.5281/zenodo.438045, STFC Scientific Computing Department’s SCARF cluster and the Sulis Tier 2 HPC platform hosted by the Scientific Computing Research Technology Platform at the University of Warwick and funded by EPSRC GrantS EP/T022108/1 and the HPC Midlands+ consortium. Via our membership of the UK’s HEC Materials Chemistry
Consortium, which is funded by EPSRC (EP/R029431), this work used the ARCHER UK National Supercomputing Service (http://www.archer.ac.uk)

\bibliographystyle{unsrtnat}

\newpage
\appendix

\section{Kernels and derivation for SPME}
\label{spme-kernel}

Below follows a derivation for the generic kernels for order $1/r^{p}$ potentials.

\paragraph{$g_{p}$}\ 

As described in \citet{essmann1995smooth} the general form of the $g$ function for $1/r^{p}$ is given as:
\begin{equation}
    g_p\brac{x} = \frac{2}{\Gamma{}\brac{\frac{p}{2}}}\int\limits_{x}^{\infty{}}s^{p-1}\exp\brac{-s^{2}}\intd{s}
\end{equation}

From \citet{GradRyz} (pp. 108--109) and \cite{essmann1995smooth} we find for $r^{-1}$ (Coulomb):
\begin{equation}
    g_{1}\brac{x} = \frac{2}{\Gamma{}\brac{\frac{1}{2}}}\int\limits_{x}^{\infty{}}\exp\brac{-s^{2}}\intd{s}
\end{equation}
\begin{equation}
    \text{where: } \int{}\exp\brac{-\beta{}x^{n}}\intd{x} = \frac{1}{2}\sqrt{\frac{\pi{}}{\beta{}}}\textrm{erf}\,\brac{\sqrt{\beta{}}x} \notag    
\end{equation}    
\begin{align}
    g_{1}\brac{x} &= \frac{2}{\sqrt{\pi{}}}\frac{\sqrt{\pi}}{2}\brac{\mathrm{erf}\,\brac{\infty} - \mathrm{erf}\,\brac{x}} \\
    g_{1}\brac{x} &= \mathrm{erfc}\,\brac{x}
\end{align}

For the general case ($1/r^{p}$) we find:
\begin{equation}
    g_{p} = \frac{2}{\Gamma{}\brac{\frac{p}{2}}}\int\limits_{x}^{\infty{}}s^{p-1}\exp\brac{-s^{2}}\intd{s}
\end{equation}

where:

\begin{align*}
    \int{}\!x^{m}\exp&\brac{\pm{}ax^{n}}\intd{x} = \\
    &\pm{}\frac{x^{m+1-n}}{na} \\
    &\mp{}\frac{m+1-n}{na}\int{}x^{m-n}\exp\brac{\pm{}ax^{n}}\intd{x}
\end{align*}

So for our simpler case:
\begin{align}
    \int{}x^{p-1}\exp&\brac{-x^{2}}\intd{x} =\nonumber\\ &-\frac{x^{p-2}}{2} \nonumber\\&+ \frac{p-2}{2}\int{}x^{p-3}\exp\brac{-x^{2}}\intd{x}
\end{align}
Which, using a series expansion gives for even powers:
\begin{equation}
    \sum\limits_{n=0}^{p/2}\, \sqbrac{\frac{x^{2n}}{n!}}\exp\brac{-x^{2}}
\end{equation}
and for odd powers:
\begin{equation}
    \sum\limits_{n=1}^{p-1} \sqbrac{\frac{2^{n}}{n!!}x^{2n-1}}\frac{\exp\brac{-x^{2}}}{\sqrt{\pi{}}} + \mathrm{erfc}\,\brac{x}
\end{equation}
where: $$n!! = \prod\limits_{j=1}^{\frac{n+1}{2}}\brac{2j-1}$$

which gives us the final form of the equation:
\begin{equation}
    g_{p}\brac{x} = 
    \begin{cases}
        \frac{2}{\Gamma{}\;\brac{\frac{p}{2}}}\sqbrac{
        \sum\limits_{n=0}^{p/2}\, \sqbrac{\frac{x^{2n}}{n!}}\exp\brac{-x^{2}}}, & p \in{} \mathbf{even} \\ 
        \frac{2}{\Gamma{}\;\brac{\frac{p}{2}}}\sqbrac{
        \sum\limits_{n=1}^{p-1} \sqbrac{\frac{2^{n}}{n!!}x^{2n-1}}\frac{\exp\brac{-x^{2}}}{\sqrt{\pi{}}} + \mathrm{erfc}\,\brac{x}}, & p \in{} \mathbf{odd}
    \end{cases}
\end{equation}
\paragraph{$f_{p}$}

\begin{equation}
    f_p\brac{x} = \frac{2x^{p-3}}{\Gamma{}\brac{\frac{p}{2}}}
    \int\limits_{x}^{\infty{}}s^{2-p}\exp\brac{-s^{2}}\intd{s}
\end{equation}

There are two simpler to solve cases, these are:

where $p = 1$:
\begin{equation}
f_{1} = \frac{1}{\sqrt{\pi{}}}\frac{\exp\,\brac{-x^{2}}}{x^{2}},
\end{equation}

and where $p$ is odd and $p > 3$
\begin{multline}
f_{p} = \\
    \frac{1}{p!\Gamma{}\;\brac{\frac{p}{2}}}
    \biggl[\sum\limits_{n=0}^{\frac{p}{2}}\brac{\brac{p-2n}!\,\brac{-1}^{n+1}x^{2n}}\exp\brac{-x^{2}} + \\
    \brac{-x^{2}}^{\frac{p-3}{2}-1}E_{i}\brac{-x^{2}}\biggr], 
\end{multline}

The general case of {$f_{p}$} is harder to evaluate. Consider

\begin{equation}
    I\brac{t,x} = \int\limits_{x}^{\infty{}}s^{t}\exp\brac{-s^{2}}\intd{s}
\end{equation}

This can be written
\begin{equation}
    I\brac{t,x} = \int\limits_{x}^{\infty{}}\sqbrac{s^{t-1}}\sqbrac{s\exp\brac{-s^{2}}}\intd{s}
\end{equation}

Which can easily be integrated by parts to give
\begin{equation}
    I\brac{t,x} = \frac{t-1}{2}I\brac{t-2,x}+\frac{1}{2}x^{t-1}\exp\brac{-x^{2}}
\end{equation}

This can be rearranged to give us an upwards recurrence
\begin{equation}
    I\brac{t+2,x} = \frac{t+1}{2}I\brac{t,x}+\frac{1}{2}x^{t+1}\exp\brac{-x^{2}}
\end{equation}

From which given I(0,x) and I(1,x) we can calculate all values of the integral for positive t. These starting integrals are straightfoward:
\begin{equation}
    I(0,x) = \frac{\sqrt{\pi}}{2} erfc(x)
\end{equation}
\begin{equation}
    I(1,x) = \frac{1}{2}\exp\brac{-x^{2}}
\end{equation}

For negative t we need a downwards recurrence. This can also be obtained by rearranging the above
\begin{equation}
    I\brac{t-2,x} = \frac{2}{t-1}\sqbrac{I\brac{t,x}-\frac{1}{2}x^{t-1}\exp^{-x^{2}}}
\end{equation}

While we can use the above for even negative t it is not suitable for odd as stepping from $t=1$ to $t=-1$ introduces a division by zero. We can solve this by noting
\begin{equation}
    I\brac{-1,x} = \frac{1}{2}E_1(x)
\end{equation}
where {$E_1\brac{x}$} is the first order exponential integral and using this as the starting point for odd negative $t$.

From $I\brac{t,x}$ $f_p$ is trivially calculated

\paragraph{$\tdbyd{g_p}{x}$}\ \\
\begin{align}
    g\brac{x} &= \frac{2}{\Gamma{}\brac{\frac{p}{2}}}\int\limits_{x}^{\infty{}}s^{p-1}\exp\brac{-s^{2}}\intd{s}\\
    g\brac{x} &= \frac{2}{\Gamma{}\brac{\frac{p}{2}}}\sqbrac{F\brac{\infty{}} - F\brac{x}} \\
    \text{where : } F\sqbrac{s} &= \int{}s^{p-1}\exp\brac{-s^{2}}\intd{s} \\
    \tdbyd{g\brac{x}}{x} &= \frac{2}{\Gamma{}\brac{\frac{p}{2}}}\sqbrac{ -F'\brac{x} } \\
    \tdbyd{g\brac{x}}{x} &= \frac{2}{\Gamma{}\brac{\frac{p}{2}}}\sqbrac{ -x^{p-1}\exp\!\brac{ -x^{2} } }
\end{align}

\paragraph{$\tdbyd{f_p}{x}$}\ \\

\begin{align}
    f\brac{x} &= \frac{2x^{p-3}}{\Gamma{}\brac{\frac{p}{2}}}\int\limits_{x}^{\infty{}}s^{2-p}\exp\brac{-s^{2}}\intd{s}\\
    f\brac{x} &= \frac{2x^{p-3}}{\Gamma{}\brac{\frac{p}{2}}}\sqbrac{F\brac{\infty{}} - F\brac{x}} \\
    \text{where : } F\sqbrac{s} &= \int{}s^{2-p}\exp\brac{-s^{2}}\intd{s} \\
    \tdbyd{f\brac{x}}{x} &= \tdbyd{}{x}\sqbrac{\frac{2x^{p-3}}{\Gamma{}\brac{\frac{p}{2}}}}\sqbrac{F\brac{\infty{}} - F\brac{x}} \nonumber\\
     &+ \frac{2x^{p-3}}{\Gamma{}\brac{\frac{p}{2}}}\tdbyd{}{x}\sqbrac{F\brac{\infty{}} - F\brac{x}} \\
     &= \frac{p-3}{x}\frac{2}{\Gamma{}\brac{\frac{p}{2}}}x^{p-3}\sqbrac{F\brac{\infty{}} - F\brac{x}} \nonumber\\
     &+ \frac{2x^{p-3}}{\Gamma{}\brac{\frac{p}{2}}}\sqbrac{-F'\brac{x}} \\
     &= \frac{p-3}{x}f\brac{x} - \frac{2x^{p-3}}{\Gamma{}\brac{\frac{p}{2}}}\sqbrac{x^{2-p}\exp\brac{-x^{2}}} \\
     &= \frac{p-3}{x}f\brac{x} - \frac{2}{x\Gamma{}\brac{\frac{p}{2}}}\exp\brac{-x^{2}}
     \label{BigProblem}
\end{align}
\textbf{N.B.} our $x$ is actually:
\begin{equation}
    x = \frac{\pi{}m}{\beta{}}    
\end{equation}
so we find that:
\begin{equation}
    \tdbyd{f\brac{x}}{m} \xrightarrow{} \tdbyd{x}{m}\tdbyd{f\brac{x}}{x}
\end{equation}
which gives:
\begin{equation}
    \frac{\pi{}}{\beta{}}\tdbyd{f\brac{x}}{x}
\end{equation}
as our actual stress kernel.

\section{Per-particle contributions for SPME}
\label{PPSPME}

This section follows the notation of the original SPME paper \cite{essmann1995smooth}. $\alpha{}$, $\beta{}$, $\gamma{}$ label the lattice
vectors. Note all is in the basis of the lattice vectors, should cartesian values be required a basis transformation is
needed at the end (see the end of the force calculation for an example in the code)

Consider

\begin{equation*}
\Omega ^{\mathit{ABC}}=\frac 1{2\pi V}\sum _{\mathbf m\neq 0}S(\mathbf m)m_{\alpha }^Am_{\beta
}^Bm_{\gamma }^Cf(\mathbf m)
\end{equation*}
where $A$, $B$, $C$ are the derivatives in $u$, $v$, $w$.

It will be shown that all of the energy, force and stresses can be expressed in this form

Given the definition of the structure factor we can write:

\begin{equation*}
\Omega ^{\mathit{ABC}}=\frac 1{2\pi V}\sum _{\mathbf m\neq 0}\sum _jq_je^{2\pi i\mathbf
m.\mathbf r_j}m_{\alpha }^Am_{\beta }^Bm_{\gamma }^Cf(\mathbf m)
\end{equation*}
Now applying the standard SPME methods. 

First start working in scaled coordinates, where the scaling is by the size
of the FFT grid we intend to use. So for a grid of size $K$\textsubscript{1}*$K$\textsubscript{2}*$K$\textsubscript{3}

\begin{equation*}
\begin{gathered}u_{\mu j}=K_{\mu }r_{\mu j}\\\Omega ^{\mathit{ABC}}=\frac 1{2\pi V}\sum _jq_j\sum _{\mathbf
m\neq 0}m_{\alpha }^Am_{\beta }^Bm_{\gamma }^Cf(\mathbf m)\prod _{\mu }e^{2\pi i\frac{m_{\mu }u_{\mu j}}{K_{\mu
}}}\end{gathered}
\end{equation*}

Now applying the SPME approximation:

\begin{equation*}
e^{2\pi i\frac{m_{\mu }}{K_{\mu }}u_{\mu j}}\approx b_{\mu }(m_{\mu })\sum _{k_{\mu }=-\infty }^{\infty }M_n(u_{\mu
j}-k_{\mu })e^{2\pi i\frac{m_{\mu }}{K_{\mu }}k_{\mu }}
\end{equation*}

Apply periodic boundary conditions:

\begin{equation*}
e^{2\pi i\frac{m_{\mu }}{K_{\mu }}u_{\mu j}}\approx b_{\mu }(m)\sum _{n_{\mu }=-\infty }^{\infty }\sum _{k_{\mu
}=0}^{K_{\mu }-1}M_n(u_{\mu j}-k_{\mu }-n_{\mu }K_{\mu })e^{2\pi i\frac{m_{\mu }}{K_{\mu }}k_{\mu }}
\end{equation*}

However we can go one step further. Differentiating A times on both sides w.r.t.  $u_{\mu{}j}$ and rearranging gives

\begin{align*}
m_{\mu }^A&e^{2\pi i\frac{m_{\mu }}{K_{\mu }}u_{\mu j}}\approx \\
&\sqbrac{\frac{K_{\mu }}{2\pi i}}^Ab_{\mu }(m) \times \\
&   \sum_{n_{\mu }=-\infty }^{\infty }
    \sum _{k_{\mu }=0}^{K_{\mu }-1}\partial ^A
        M_n(u_{\mu j}-k_{\mu }-n_{\mu }K_{\mu})/
        \partial u_{\mu j}e^{2\pi i\frac{m_{\mu }}{K_{\mu }}k_{\mu }}
\end{align*}
The properties of cardinal B splines means this is easy to evaluate as

\begin{equation*}
\frac d{\mathit{du}}M_n(u)=M_{n-1}(u)-M_{n-1}(u-1)
\end{equation*}
and since the splines are evaluated by recursion we already have all the terms we need already.

Inserting this into the above gives

\begin{align*}
\Omega ^{\mathit{ABC}} =&
\frac 1{2\pi V}
\sum _j\sum _{n_1,n_2,n_3=-\infty }^{\infty }\sum _{k_1}^{K_1-1}\sum_{k_2}^{K_2-1}\sum _{k_3}^{K_3-1} \left[
Q_j^{\mathit{ABC}}(\mathbf k,\mathbf n) \times \right. \\
& \left. \sum _{\mathbf m\neq0}\left[\prod _{\mu }b_{\mu }(\mathbf m)e^{2\pi i\frac{m_{\mu }k_{\mu }}{K_{\mu }}}\right]f(\mathbf
m) \right]
\end{align*}

\begin{align*}
Q_j^{\mathit{ABC}}(\mathbf k,\mathbf n) = 
q_j&\left[\frac{K_{\alpha }}{2\pi i}\right]^A\frac{\partial^A}{\partial u_{\alpha j}^A}M_n(u_{\alpha j}-k_{\alpha }-n_{\alpha }K_{\alpha })\times \\
&\left[\frac{K_{\beta }}{2\pi i}\right]^B\frac{\partial ^B}{\partial u_{\beta j}^B}M_n(u_{\beta j}-k_{\beta }-n_{\beta }K_{\beta }) \times \\
&\left[\frac{K_{\gamma }}{2\pi i}\right]^C\frac{\partial ^C}{\partial u_{\gamma j}^C}M_n(u_{\gamma j}-k_{\gamma
}-n_{\gamma }K_{\gamma })
\end{align*}

Now the sum over $m$ is almost a discrete Fourier transform. The problems are missing out zero and the infinite
range.

We can make it one if

\begin{enumerate}
\item We define f(\textbf{0})=0 -- which it will be for a neutral cell 
\item We only sum out to $K$\textsubscript{$\mu $}{}-1. This is OK if f(\textbf{m}) has decayed to negligible values by
the edge of the grid. This should be the case if we have converged the calculation carefully -- ultimately the decaying
potential term will kill off the function provided we have chosen the grid size and Ewald parameter correctly
\end{enumerate}

Thus we can define a per particle contribution as
\begin{align}
\Omega ^{\mathit{ABC}}&=\sum _j\omega _j^{\mathit{ABC}}\\\omega _j^{\mathit{ABC}}&=\frac 1{2\pi V}\sum
_{n_1n_2n_3=-\infty }^{\infty }\sum _{k_1}^{K_1-1}\sum _{k_2}^{K_2-1}\sum
_{k_3}^{K_3-1}Q_j^{\mathit{ABC}}(\mathbf k,\mathbf n)\sum _{\mathbf m\neq 0}\left[\prod _{\mu
}b_{\mu }(\mathbf m)e^{2\pi i\frac{m_{\mu }k_{\mu }}{K_{\mu }}}\right]f(\mathbf
m)\\Q_j^{\mathit{ABC}}(\mathbf k,\mathbf n)&=q_j\left[\frac{K_{\alpha }}{2\pi i}\right]^A\frac{\partial
^A}{\partial u_{\alpha j}^A}M_n(u_{\alpha j}-k_{\alpha }-n_{\alpha }K_{\alpha })\times \nonumber\\&\left[\frac{K_{\beta }}{2\pi
i}\right]^B\frac{\partial ^B}{\partial u_{\beta j}^B}M_n(u_{\beta j}-k_{\beta }-n_{\beta }K_{\beta })\times
\nonumber\\&\left[\frac{K_{\gamma }}{2\pi i}\right]^C\frac{\partial ^C}{\partial u_{\gamma j}^C}M_n(u_{\gamma j}-k_{\gamma
}-n_{\gamma }K_{\gamma })
\end{align}
\newpage
\section{Redefinition of control files}
\label{new-control}
\paragraph{Old Style}

\begin{verbatim}
P-cresol 340

temperature  340
ensemble nve      1.0
pressure          0.000    
steps             10200000
multiple          1
print             2000
stack             100
stats             2000
trajectory        200000 2000 0
timestep          0.0005
cutoff            10.0
delr              0.5
rvdw cutoff	        10.000E+00                     
ewald precision	        1.0000E-06                     
cap		        1.0000E+04                     
shake tolerance	        1.0000E-05                     
quaternion tolerance    1.0000E-05                    	                                       
job time	        2.5000E+05
close time	        1.0000E+02
finish
\end{verbatim}

\newpage

\paragraph{New Style}

\begin{verbatim}
title P-cresol 340

io_file_config CONFIG
io_file_field FIELD
io_file_statis STATIS
io_file_revive REVIVE
io_file_revcon REVCON
io_file_cor COR
io_file_currents CURRENTS

print_frequency 2000 steps
stats_frequency 2000 steps
stack_size 100 steps

vdw_cutoff 10.0 ang
padding 0.125 ang
cutoff 10.0 ang
coul_method spme
spme_precision 1e-06

ensemble nve
temperature 340.0 K
pressure_hydrostatic 0.0 katm
timestep 0.0005 ps

time_run 10200000 steps
time_equilibration 0 steps
restart clean

traj_calculate ON
traj_start 200000 steps
traj_interval 2000 steps
traj_key pos

initial_minimum_separation 0.0 ang
equilibration_force_cap 10000.0 k_B.temp/ang
shake_tolerance 1e-05 ang
\end{verbatim}
\newpage
\section{Implementation of general correlations}\label{app:cor}
In order to support arbitrary correlations in a modular and extensible manner, we separate the code into two components. First implementing the multi-tau algorithm \cite{Ramirez2010Time}, and second supplying an interface to submit data for correlation at runtime.

For the first, Algorithm \ref{alg:update} details updating a correlator with newly observed data $x$ and $y$. In this listing the correlator structure includes rolling shift arrays for the data to be stored in, accumulators, and correlation values, all index by $b_{i}$ the block index in the hierarchical data structure. Every $m$ steps the algorithm proceeds to the next block by averaging the accumulated values of $x$ and $y$ that have previously been submitted. This step results in the increasing granularity when $b_{i}> 1$ if $m > 1$. Every $p$ steps the shift arrays will again update from the start. The values of $x$ and $y$ may be real or complex (in which case a conjugate must be taken). In order to read out the correlation values, the implicit time lags due to any coarse graining need to be unwrapped as shown in Algorithm \ref{alg:readout}. 

During a simulation multiple correlators are spawned, identified with correlating particular observables. As the simulation progresses, whenever a correlation must be updated the get method for the observables it tracks are called and the data submitted with $b_{i}=1$ (the first block).
 
\begin{algorithm}
\SetKwInOut{Input}{input}
\label{alg:update}
\caption{Updating a correlator}
    \Input{$x$: first value correlated, $y$: second value correlated, $b_{i}$: the correlator block to update, cor: correlator data to update.}
    \If{$b_{i} > $ cor.blocks}{\Return}
    \text{cor.blocksUsed} $\gets \text{max}(b_{i}, \text{cor.blocksUsed})$\\
    cor.shift[$b_{i}$, cor.shift[$b_{i}$], :] $\gets x, y$\\
    cor.validShift[$b_{i}$] $\gets$ max(cor.shift[$b_{i}$], 1)\\
    cor.accumulator[$b_{i}$, :] $\gets$ cor.accumulate[$b_{i}$, :] $+ x, y$\\
    cor.accumulated[$b_{i}$] $\gets$ cor.accumulated[$b_{i}$]+1\\
    \If{cor.accumulated[$b_{i}$] $>$ \text{cor.m}}{
      $u \gets$ cor.accumulator[$b_{i}$, 1]$/$cor.accumulated[$b_{i}$]\\
      $v \gets$ cor.accumulator$b_{i}$, 2]$/$cor.accumulated[$b_{i}$]\\
      Update($u$, $v$, $b_{i}+1$, cor)\\
      cor.accumulator[$b_{i}$, :] $\gets 0$; cor.accumulated[$b_{i}$, :] $\gets 0$\\
    }
    $i \gets$ cor.shift[$b_{i}$]\\
    \eIf{$i == 1$}
    {
        $j \gets i$\\
        \For {$n \gets 1, \text{cor.validShifts}[b_{i}]$}
        {
            cor.values[$b_{i}, n$] $\gets$ cor.values[$b_{i}, n$]+cor.shift[$b_{i}, n, 1$]*\text{Conj}(cor.shift[$b_{i}, j, 2$])\\
            cor.count[$b_{i}, n$] $\gets$ cor.count[$b_{i}, n$] $+1$\\
            $j \gets j-1$\\
            \If {$j <= 0$}{$j\gets j + p$}
        }
    }
    {
        $j \gets i - \lfloor\text{cor.p}/\text{cor.m}\rfloor+1$\\
        \For {$n \gets \lfloor\text{cor.p}/\text{cor.m}\rfloor+1, \text{cor.p}$}
        {
            \If {$j <= 0$}{$j\gets j + \text{cor.p}$}
            cor.values[$b_{i}, n$] $\gets$ cor.values[$b_{i}, n$]+cor.shift[$b_{i}, n, 1$]*\text{Conj}(cor.shift[$b_{i}, j, 2$])\\
            cor.count[$b_{i}, n$] $\gets$ cor.count[$b_{i}, n$] $+1$\\
            $j \gets j-1$\\
        }
    }
    cor.shift[$b_{i}$] $\gets$ cor.shift[$b_{i}$]$+1$\\
    \If {cor.shift[$b_{i}$] $>$ $p$}{cor.shift[$b_{i}$] $\gets 1$}
\end{algorithm}

\begin{algorithm}
\SetKwInOut{Input}{input}
\SetKwInOut{Output}{output}
\label{alg:readout}
\caption{Readout of a correlation}
    \Input{cor: correlator to readout.}
    \Output{values: the correlation values, times: the time lags between correlation values.}
    $\tau \gets 1$\\
    \For {$i \gets 1,$ cor.p}
    {
      \If {cor.count[$1, i$] $> 0$}{values[$\tau$]$\gets$ values[$\tau$] $+$ cor.values[$1,i$]/cor.count[$1,i$]\\ times[$\tau$] $\gets i-1$\\ $\tau\gets \tau+1$}
    }
    \For {$k \gets 1, \text{cor.blocksUsed}$}
    {
        \For {$i \gets \lfloor\text{cor.p}/\text{cor.m}\rfloor+1,$ cor.p}
        {
          \If {cor.count[$k, i$] $> 0$}{values[$\tau$]$\gets$ values[$\tau$] $+$ cor.values[$k,i$]/cor.count[$k,i$]\\ times[$\tau$] $\gets (i-1)*m^{k-1}$\\ $\tau \gets \tau + 1$}
        }
    }
\end{algorithm}

For the second part, the interface calls for a pair of abstract \texttt{Observable} types to be supplied which each implement a get method. This method obtains the data to be correlated and acts as a kernel. By designing the module in this way a single loop over arbitrary pairs of \texttt{Observable} types may be completed to compute all correlations, without the need for additional selection logic. The specific functionality is implemented in types deriving from \texttt{Observable} as opposed to including this logic in the statistics collection subroutine itself. For some types this is a relatively trivial task. For example obtaining the velocity of a particular atom. Some a little more complex such as selecting one of the six current types which are also separated by atomic species and {\bf k}-point. In all cases however, computation is separated from obtaining the data.

\section{Benchmark data}\label{benchmark-data}
\begin{table}
  \centering
  \renewcommand{\arraystretch}{1.2}
  \begin{tabular}{|c||c|c|c|c|c|c|}
    \hline
    Method & \multicolumn{3}{c|}{On-the-fly} & \multicolumn{3}{c|}{Trajectory Saving} \\
    \cline{1-7}
    MPI Cores & 128 & 640 & 1280 & 128 & 640 & 1280\\
    \cline{1-7}
    Baseline (s) & 123.19 & 66.14 & 75.40 & 123.19 & 66.14 & 75.40 \\
    \hhline{=======}
    Sampling interval &\multicolumn{6}{c|}{Wall time (s)} \\ \hline
    1 & 617.24 & 190.31 & 145.13 & 1072.34 & 3896.31 & 4182.64 \\ \hline
    2 & 401.74 & 129.71 & 103.83 & 597.86 & 2464.96 & 2383.92 \\ \hline
    4 & 312.15 & 99.45 & 90.40 & 353.98 & 1623.54 & 1280.41 \\ \hline
    8 & 268.33 & 87.94 & 82.55 & 238.28 & 1305.37 & 548.13 \\ \hline
    16 & 245.25 & 81.08 & 84.52 & 179.88 & 689.58 & 650.48 \\ \hline
    32 & 235.49 & 80.96 & 81.70 & 147.60 & 374.37 & 457.52 \\ \hline
  \end{tabular}
  \caption{Absolute wall times for the strong scaling benchmarks in figure \ref{fig:vaf_strong} for the on-the-fly and trajectory saving methods. The baseline walls times indicate simulations performed with neither trajectory saving or on-the-fly correlations.}
  \label{strong-sim}
\end{table}

\begin{table}
  \centering
  \renewcommand{\arraystretch}{1.2}
  \begin{tabular}{|c||c|c|c|c|c|c|}
    \hline
    Method & \multicolumn{3}{c|}{Replay} & \multicolumn{3}{c|}{Python FFT} \\
    \cline{1-7}
    MPI Cores & 128 & 640 & 1280 & 128 & 640 & 1280\\
    \cline{1-7}
    Baseline (s) & 123.19 & 66.14 & 75.40 & 123.19 & 66.14 & 75.40 \\
    \hhline{=======}
    Sampling interval &\multicolumn{6}{c|}{Wall time (s)} \\ \hline
    1 & 807.66 & 1027.33 & 1197.84 & 2367.95 & 2513.56 & 2815.13 \\ \hline
    2 & 351.52 & 505.57 & 604.71 & 1182.09 & 1241.04 & 1395.91 \\ \hline
    4 & 123.73 & 264.06 & 405.30 & 599.24 & 656.39 & 720.67 \\ \hline
    8 & 48.62 & 154.71 & 213.92 & 295.79 & 320.55 & 365.25 \\ \hline
    16 & 32.87 & 79.75 & 124.14 & 148.95 & 178.81 & 193.01 \\ \hline
    32 & 23.63 & 55.94 & 93.88 & 78.74 & 91.49 & 110.22 \\ \hline
  \end{tabular}
  \caption{Absolute wall times for the strong scaling benchmarks in figure \ref{fig:vaf_strong} for post-processing methods. The baseline walls times indicate simulations performed with neither trajectory saving or on-the-fly correlations.}
  \label{strong-post}
\end{table}

\begin{table}
  \centering
  \renewcommand{\arraystretch}{1.2}
  \begin{tabular}{|c||c|c|c|c|c|}
    \hline
    Atoms & 8000 & 27000 & 64000 & 125000 & 512000 \\
    \cline{1-6}
    Baseline (s) & 16.00 & 16.43 & 20.34 & 26.86 & 49.89 \\
    \hhline{======}
    Sampling interval &\multicolumn{5}{c|}{Wall time (s)} \\ \hline
    1 & 34.14 & 54.03 & 61.64 & 70.61 & 114.15 \\ \hline
    2 & 22.36 & 32.65 & 31.62 & 45.16 & 80.36 \\ \hline
    4 & 17.79 & 24.81 & 24.65 & 34.30 & 68.73 \\ \hline
    8 & 17.52 & 21.22 & 21.31 & 30.97 & 62.61 \\ \hline
    16 & 17.12 & 18.10 & 20.26 & 28.62 & 59.04 \\ \hline
    32 & 15.98 & 17.87 & 20.02 & 29.65 & 57.70 \\ \hline
  \end{tabular}
  \caption{Absolute wall times for the on-the-fly method weak scaling benchmarks in figure \ref{fig:vaf_weak}. The baseline walls times indicate simulations performed with neither trajectory saving or on-the-fly correlations.}
  \label{weak-otf}
\end{table}

\begin{table}
  \centering
  \renewcommand{\arraystretch}{1.2}
  \begin{tabular}{|c||c|c|c|c|c|}
    \hline
    Atoms & 8000 & 27000 & 64000 & 125000 & 512000 \\
    \cline{1-6}
    Baseline (s) & 16.00 & 16.43 & 20.34 & 26.86 & 49.89 \\
    \hhline{======}
    Sampling interval &\multicolumn{5}{c|}{Wall time (s)} \\ \hline
    1 & 126.32 & 116.81 & 530.03 & 437.78 & 2701.92 \\ \hline
    2 & 63.28 & 68.59 & 147.26 & 197.84 & 1802.02 \\ \hline
    4 & 39.19 & 39.57 & 69.90 & 130.18 & 1356.89 \\ \hline
    8 & 26.89 & 35.86 & 48.43 & 72.15 & 997.88 \\ \hline
    16 & 21.32 & 20.06 & 33.32 & 52.30 & 542.28 \\ \hline
    32 & 16.26 & 18.17 & 25.22 & 36.30 & 322.02 \\ \hline
  \end{tabular}
  \caption{Absolute wall times for the trajectory saving method weak scaling benchmarks in figure \ref{fig:vaf_weak}. The baseline walls times indicate simulations performed with neither trajectory saving or on-the-fly correlations.}
  \label{weak-save}
\end{table}

\begin{table}
  \centering
  \renewcommand{\arraystretch}{1.2}
  \begin{tabular}{|c||c|c|c|c|c|}
    \hline
    Atoms & 8000 & 27000 & 64000 & 125000 & 512000 \\
    \cline{1-6}
    Baseline (s) & 16.00 & 16.43 & 20.34 & 26.86 & 49.89 \\
    \hhline{======}
    Sampling interval &\multicolumn{5}{c|}{Wall time (s)} \\ \hline
    1 & 29.69 & 70.58 & 73.13 & 258.03 & 804.66 \\ \hline
    2 & 9.40 & 24.31 & 20.64 & 63.96 & 401.23 \\ \hline
    4 & 4.98 & 13.34 & 8.90 & 33.96 & 215.81 \\ \hline
    8 & 2.65 & 5.57 & 6.60 & 26.34 & 116.58 \\ \hline
    16 & 1.75 & 5.01 & 5.04 & 16.57 & 48.43 \\ \hline
    32 & 1.44 & 2.87 & 4.06 & 15.30 & 45.05 \\ \hline
  \end{tabular}
  \caption{Absolute wall times for the replay post-processing method weak scaling benchmarks in figure \ref{fig:vaf_weak}. The baseline walls times indicate simulations performed with neither trajectory saving or on-the-fly correlations.}
  \label{weak-replay}
\end{table}

\begin{table}
  \centering
  \renewcommand{\arraystretch}{1.2}
  \begin{tabular}{|c||c|c|c|c|c|}
    \hline
    Atoms & 8000 & 27000 & 64000 & 125000 & 512000 \\
    \cline{1-6}
    Baseline (s) & 16.00 & 16.43 & 20.34 & 26.86 & 49.89 \\
    \hhline{======}
    Sampling interval &\multicolumn{5}{c|}{Wall time (s)} \\ \hline
    1 & 20.08 & 65.65 & 145.54 & 291.43 & 1242.89 \\ \hline
    2 & 10.27 & 32.97 & 73.62 & 144.26 & 632.86 \\ \hline
    4 & 5.23 & 16.80 & 37.07 & 75.26 & 314.30 \\ \hline
    8 & 2.77 & 8.26 & 18.95 & 37.75 & 160.83 \\ \hline
    16 & 1.30 & 4.30 & 9.71 & 19.20 & 90.36 \\ \hline
    32 & 0.75 &  2.27 & 5.06 & 10.02 & 45.99 \\ \hline
  \end{tabular}
  \caption{Absolute wall times for the Python FFT post-processing method weak scaling benchmarks in figure \ref{fig:vaf_weak}. The baseline walls times indicate simulations performed with neither trajectory saving or on-the-fly correlations.}
  \label{weak-python}
\end{table}
\end{document}